\documentclass[fleqn,usenatbib]{mnras}

\usepackage{newtxtext,newtxmath}
\usepackage[T1]{fontenc}

\DeclareRobustCommand{\VAN}[3]{#2}
\let\VANthebibliography\thebibliography
\def\thebibliography{\DeclareRobustCommand{\VAN}[3]{##3}\VANthebibliography}

\usepackage{graphicx}	% Including figure files
\usepackage{amsmath}	% Advanced maths commands

\newcommand{\MSUN}{{\rm M}_{\sun}}
\newcommand{\MHOST}{M_{\rm 200c, host}}
\newcommand{\MSTAR}{M_{\rm *}}

\title[Cumulative SFHs of TNG50 dwarfs: environment and quenching]{The cumulative star-formation histories of dwarf galaxies with TNG50. I: Environment-driven diversity and connection to quenching}

\author[G. D. Joshi et al.]{Gandhali D. Joshi,$^{1}$\thanks{E-mail: joshi@mpia.de}
Annalisa Pillepich,$^{1}$
Dylan Nelson,$^{2}$
Elad Zinger,$^{1,3}$
Federico Marinacci,$^{4}$ \newauthor
Volker Springel,$^{5}$
Mark Vogelsberger,$^{6}$
and Lars Hernquist$^{7}$
\\ \\
$^{1}$Max-Planck-Institut f{\"u}r Astronomie, K{\"o}nigstuhl 17, 69117 Heidelberg, Germany\\
$^{2}$Universit\"{a}t Heidelberg, Zentrum f\"{u}r Astronomie, Institut f\"{u}r theoretische Astrophysik, Albert-Ueberle-Str. 2, 69120 Heidelberg, Germany\\
$^{3}$Centre for Astrophysics and Planetary Science, Racah Institute of Physics, The Hebrew University, Jerusalem 91904, Israel\\
$^{4}$Department of Physics and Astronomy ``Augusto Righi'', Universit{\`a} di Bologna, Via Gobetti 93/2, 40129 Bologna, Italy\\
$^{5}$Max Planck Institut f{\"u}r Astrophysik, Karl-Schwarzschild-Straße 1, 85748 Garching bei M{\"u}nchen, Germany\\
$^{6}$Kavli Institute for Astrophysics and Space Research, Massachusetts Institute of Technology, Cambridge, MA 02139, USA \\
$^{7}$Institute for Theory and Computation, Harvard-Smithsonian Center for Astrophysics, 60 Garden St., MS-51, Cambridge, MA 02138, USA
}

\date{Accepted XXX. Received YYY; in original form ZZZ}

\pubyear{2020}

\begin{document}
\label{firstpage}
\pagerange{\pageref{firstpage}--\pageref{lastpage}}
\maketitle

\begin{abstract}
We present the cumulative star-formation histories (SFHs) of $>15000$ dwarf galaxies ($\MSTAR=10^{7-10}\MSUN$) simulated with the TNG50 run of the IllustrisTNG suite across a vast range of environments. We show that the key factors determining the dwarfs' SFHs are their status as central or satellite and their stellar mass, with centrals and more massive dwarfs assembling their stellar mass at later times on average compared to satellites and lower mass dwarfs. The satellites in our sample (in hosts of total mass $\MHOST=10^{12-14.3}\MSUN$) assembled 90 per cent of their $z=0$ stellar mass $\sim7.0_{-5.5}^{+3.3}$ Gyr ago, on average and within the $10^{\rm th}-90^{\rm th}$ percentiles, while the centrals did so only $\sim1.0_{-0.5}^{+4.0}$ Gyr ago. TNG50 predicts a large diversity in SFHs for both centrals and satellites, so that the stacked cumulative SFHs are representative of the TNG50 dwarf populations only in an average sense and individual dwarfs can have significantly different cumulative SFHs.
In the case of the satellites, dwarfs with the highest stellar mass to host mass ratios have the latest stellar mass assembly. Conversely, satellites at fixed stellar and host halo mass found closer to the cluster centre or accreted at earlier times show significantly earlier stellar mass assembly. These trends, as well as the shapes of the SFHs themselves, are a manifestation of the varying proportions within a given subsample of quenched vs. star-forming galaxies, which exhibit markedly distinct SFH shapes. We also find a subtle effect whereby satellite dwarfs in the most massive hosts at $z=0$ are more efficient, i.e. have higher SFRs, at early times, well \emph{before} final infall into their $z=0$ host, compared to a control sample of centrals mass-matched at the time of accretion. This suggests that the large-scale environment can have a mild effect even on \emph{future} satellites by providing the conditions for enhanced SF at early epochs. Our results are useful theoretical predictions for comparison to future resolved-stellar-population observations.
\end{abstract}

%galaxies: star formation -- 
%%%%%%%%%%%%%%%%%%%%%%%%%%%%%%%%%%%%%%%%%%%%%%%%%%

%%%%%%%%%%%%%%%%% BODY OF PAPER %%%%%%%%%%%%%%%%%%

\section{Introduction}

Galaxies in the Universe encompass several orders of magnitude in stellar mass, from $\MSTAR\gtrsim10^{12}\MSUN$ brightest cluster galaxies (BCGs) at the centres of massive galaxy clusters, to Milky Way (MW)-mass galaxies containing $10^{10.5-11}\MSUN$ of stars, down to dwarf galaxies with stellar masses of $\MSTAR\sim10^{6-10}\MSUN$, and even smaller. This vast range in mass is thought to be accompanied by an incredible diversity of formation and growth histories, of physical processes that regulate the stellar and gas content and properties of the galaxies, and of their response to their local and large-scale environments. 

Dwarf galaxies in particular ($\MSTAR\lesssim10^{10}\MSUN$) are an interesting mass regime in which to study galaxy formation and evolution -- their shallower potentials and lower masses make such galaxies especially susceptible to environmental effects, even though, in the field population, such galaxies may actually be relatively more dark-matter dominated than MW-mass galaxies (e.g. see fig. 14 from \citealt{Behroozi2013} for a compilation of observational results). Because of the nature of the hierarchical growth of structures, lower-mass galaxies can be hosted by a wider range of host halo masses than more massive ones. Furthermore, at such low masses, these dwarf galaxies are not likely to be significantly affected by feedback from active galactic nuclei (AGN), thereby allowing us to examine stellar feedback models independent from any uncertainties in AGN physics. Hence, dwarf galaxies offer unique insights into the various processes that govern galaxy evolution in the Universe.

The star-formation histories (SFHs) of galaxies directly provide us with a timeline of how they assembled their present-day stellar mass. The SFHs also contain signatures of various physical phenomena that dwarfs may have experienced, such as gas accretion and cooling, stellar and AGN feedback, mergers, in addition to environmental processes that result in removal of gas and even stellar material from the galaxy and, perhaps, quenching of star formation. Therefore, studying the SFHs of dwarfs can give us insight into their evolutionary histories, including the impact of different environments.

Dwarf galaxies are the most abundant galaxies in the Universe \citep[e.g. see galaxy stellar mass functions from][]{Panter2007,Baldry2008,Li2009,Pozzetti2010,Baldry2012,Moustakas2013,Muzzin2013}. However, their lower masses and therefore lower luminosities also make them much more difficult to observe compared to more massive galaxies, especially at higher redshifts. This in turn makes it difficult to accurately reconstruct the SFHs for a large number of dwarf galaxies. For galaxies where resolved stellar populations can be observed, SFHs can be reconstructed directly from colour-magnitude diagrams (CMDs) of individual stars, through the isolation of multiple single-age stellar populations by finding the oldest main-sequence turn-off stars \citep[e.g.][and references therein]{Dolphin2002,Weisz2011,Brown2014,Weisz2014,Gallart2015,Skillman2017,Cignoni2018,Sacchi2018,Cignoni2019,Weisz2019,Rusakov2020}. Furthermore, uncertainties in the measured absolute SFHs can be circumvented by constructing \emph{cumulative} SFHs, i.e. in the form of total stellar mass assembled up to a given time rather than an instantaneous SFR at that time, as is now common praxis in observational studies of the local Universe. Moreover, in order to stack the SFHs of several galaxies of potentially different stellar masses at $z=0$, it is standard practice to first normalize the cumulative SFHs by the final stellar mass of each galaxy.

\citet{Weisz2011}, hereafter referred to as DW11, provided uniformly measured absolute and cumulative SFHs for 60 dwarfs within a mass range of $\MSTAR\sim10^{5-11}\MSUN$ in the nearby Universe ($D\leq4$ Mpc) from the ACS Nearby Galaxy Survey Treasury (ANGST), that are found outside the Local Group (LG). By comparing the SFHs of dwarfs of various morphological types (based on morphological type $T$ including dwarf spheroidals; dSphs, spirals, dSpirals and irregulars dIs as well as transition dwarfs, dTrans, and tidal dwarfs, dTidals), they showed that although there was a large diversity in the individual SFHs, the mean SFHs of dwarfs of different morphologies were remarkably similar, with most having built up the bulk of their stellar mass at $z\gtrsim1$. Any differences between the morphological types were only seen within the last few Gyr. The dwarfs also displayed a morphology-density relation, leading the authors to conclude that environment is an important factor in transforming galaxies from gas-rich to gas-poor. Furthermore, by comparing ANGST galaxies with LG dwarfs in \citet{Weisz2011b}, the authors showed that the LG and ANGST dwarfs had similar SFHs and morphology-density relations, thereby concluding that the LG dwarfs were representative of the dwarf galaxies in the local Universe. Within the LG, \citet{Weisz2014}, hereafter referred to as DW14, also reconstructed cumulative SFHs for 40 dwarfs within a mass range of $\MSTAR\sim10^{4-8}\MSUN$: they found a stellar mass dependence such that the least massive dwarfs in their sample ($\MSTAR<10^{5}\MSUN$) formed $>80$ per cent of their stellar mass at $z>2$, whereas the more massive dwarfs had produced only $\sim30$ per cent of their stellar mass by the same epoch. Similarly, \citet{Gallart2015} suggested that the diverse SFHs of LG dwarfs can be broadly classified into two types: fast dwarfs that had an early, short period of vigorous star formation, and slow dwarfs that built only a small fraction of their stellar mass at early times and continued to add to it up to present day, with the latter being found preferentially in lower density environments. However, they argue that these two types do not necessarily correlate with specific morphological types. Most recently, using deeper data than previously available, \citet{Weisz2019} compared the SFHs of MW and M31 dwarf satellites and found significant differences, with a portion of the MW dwarf population that is more luminous having quenched more recently ($<3$ Gyr ago), compared to the more uniform quenching times (3-6 Gyr) of the M31 dwarfs, indicating the influence of environment. The majority of observational results on dwarf SFHs are currently restricted to the local Universe. However, there are some notable studies that have explored dwarfs in more distant systems, e.g. from Legacy ExtraGalactic UV Survey \citep[LEGUS;][]{Cignoni2018,Sacchi2018,Cignoni2019}. However, these observations mainly focus on the most recent SFHs (within the last few 100 Myr) of the galaxies.

In simulations, the process of measuring SFHs is made simple by the fact that we can access the entire history of a galaxy and track individual stellar particles to the time of their birth. Although dwarf galaxies can be simulated with quite good resolution, doing so within a large range of environments or in a full-cosmological context can become computationally restrictive. Therefore, previous studies of dwarf SFHs with simulations have mainly focused on dwarfs in lower-density environments, typically involving MW/M31-mass or LG-like systems with a population of dwarfs in their vicinities \citep[e.g.][]{Wetzel2016,Buck2019,GarrisonKimmel2019,Graus2019,Digby2019} or less massive haloes \citep[e.g.][]{Onorbe2015,Fitts2017,Wright2019,Mattia2020}. Note that here, we only discuss simulations of dwarfs within MW-mass or more massive hosts and not those of isolated dwarfs.

For example, \citet[][hereafter referred to as AW16]{Wetzel2016} and \citet{Buck2019}, with the Latte (based on the \textsc{fire-2} model, \citealt{Hopkins2018}) and NIHAO-UHD (based on the \textsc{gasoline2} model, \citealt{Wadsley2017}) suites of simulations, respectively, of a handful of MW-like hosts, were able to not only reproduce the abundance of observed satellites, but also to recover the wide range of observed SFHs. Using a suite of zoom-in cosmological simulations of isolated dwarfs of mass $\MSTAR\sim10^{4.5-7.5}\MSUN$ run with the \textsc{gasoline} code \citep{Wadsley2004}, \citet{Wright2019} showed that while nearly half of their dwarf population ceased star formation after the epoch of reionization, a significant fraction \emph{restarted} star-formation at later times through interaction with gas in the inter-galactic medium (IGM). \citet{Digby2019} analysed the results of the Auriga \citep{Grand2017} and APOSTLE \citep{Fattahi2016,Sawala2016} suites of simulations of MW-like and LG-like systems, respectively, and found that lower mass dwarfs ($\MSTAR=10^{5-6}\MSUN$) had steadily declining SFHs (i.e. they had a higher proportion of older stars compared to younger ones), while the opposite trend was seen for more massive dwarfs ($\MSTAR=10^{7-9}\MSUN$); intermediate mass dwarfs were shown to have approximately constant SFHs (i.e. equal proportions of old, intermediate and young stars). Additionally, they found that in contrast to field galaxies, the majority of satellite dwarfs had significantly suppressed SFRs in the last $\sim5$~Gyr and that there was little correlation between their SFHs and their distance to their host central galaxies. \citet[][hereafter referred to as GK19]{GarrisonKimmel2019} analysed one of the most extensive samples of simulated dwarf galaxies with masses $\MSTAR=10^{5-9}$ from the \textsc{fire-2} project and were able to reproduce several key observed trends: more massive dwarfs had more extended SFHs and quenched at later times; satellites had more truncated SFHs compared to `near field' dwarfs in the vicinity of LG-like or isolated MW-like systems as well as truly isolated `field' dwarfs; moreover, the `field' dwarfs had more extended SFHs compared to `near field' dwarfs around the MW-like systems, although this difference was less clear around the LG-like systems. The correlation between higher galaxy stellar mass and later assembly was also found by \cite{Simpson2018} using the results of the Auriga simulations.

While these and other prior studies have shown the rich diversity of both observed and modeled dwarf SFHs and have provided important insights, they have mostly focused on MW-like and LG-like environments and on a limited number of dwarfs, a few tens each at most. Furthermore, none of the galaxy formation models underlying the aforementioned simulation studies have been robustly tested and verified against observations of dwarfs in more dense environments, such as those of groups and clusters of galaxies like the neighbouring Fornax or Virgo clusters.

In this study, we make use of the TNG50 run of the IllustrisTNG suite of simulations\footnote{\url{https://www.tng-project.org/}} (hereafter TNG) to quantify the cumulative SFHs of thousands of $\MSTAR=10^{7-10}\MSUN$ dwarf galaxies at $z=0$ across an unprecedented range of host environments. The TNG50 simulation is an ideal dataset for this endeavour due to its combination of good resolution and sufficiently large volume, so that it returns not only a population of tens of thousands of galaxies, both satellites and centrals, but it also samples hosts as massive as Virgo (i.e. total host mass of $\sim10^{14}\MSUN$). Moreover, the underlying TNG numerical model has been shown to produce results that are consistent with a wide variety of available observational constraints. For example, in relation to the effects of environment on galaxy evolution and quoting only analyses where careful mocks of either the observable, the sample selection, or both, have been applied for the comparison to observational data, it has been shown that the TNG fractions of quenched galaxies are within $<10-20$ percentage points from the observed values at $z=0$ both at the high-mass end and in the most massive hosts \citep{Donnari2020b}, as well as for low-mass isolated galaxies \citep{Dickey2020}. In particular, in \citet{Donnari2020b} it has been shown that the halocentric-distance-dependent quenched fractions of TNG satellites in the $10^{9.7-10.5}\MSUN$ stellar mass range in intermediate group-mass hosts ($10^{13-14}\MSUN$) are in strikingly good agreement with SDSS results. The colors, stellar ages, and stellar metallicities of $\gtrsim10^9\MSUN$ TNG galaxies are either in excellent quantitative agreement or in the ball park of SDSS constraints \citep{TNGNelson2018}; and the deficit in gas content of $\gtrsim10^9\MSUN$ satellite galaxies in comparison to field analogues at low redshifts is consistent with observations of both atomic and molecular hydrogen \citep{Stevens2019,Stevens2020}.

In this analysis, we aim to quantify the impact of various environmental factors, in addition to galaxy stellar mass, in determining the SFHs of dwarf galaxies. We also want to recast the concepts of environmental quenching and time since quenching -- when this occurs -- under the formalism of the cumulative SFHs and of the stellar assembly times. In doing so, we put forward theoretical predictions and motivations for future resolved-stellar-population observations of the local Universe with upcoming telescopes like the JWST, the WFIRST/Roman, the LSST/Rubin and, further down the line, the ELT with MICADO.

The remainder of the paper is organized as follows. In Section~\ref{sec:methods}, we lay out the details of the simulation and the measurement of the galaxies' SFHs and other properties. In Section~\ref{sec:TNG50dwarfs}, we present the sample selection and the main properties of TNG50 dwarfs. The cumulative SFHs for the entire sample, satellites and centrals, and the dependence on stellar mass are analysed in Section~\ref{sec:SFHs}, including a quantification of their diversity. The role of environment on satellites, and large-scale environments on the entire dwarf sample, is explored in Section~\ref{sec:environmentalFactors}. Finally, we discuss our results and the connection to quenching in Section~\ref{sec:discussion} and summarize our findings in Section~\ref{sec:conclusions}.

%-------------------------------------------------
\section{Methods} \label{sec:methods}

\subsection{The TNG50 simulation}
We use the results from the IllustrisTNG project \citep[][hereafter referred to as TNG]{TNGMarinacci2018,TNGNaiman2018,TNGNelson2018,TNGPillepich2018,TNGSpringel2018}, which is a set of cosmological magnetohydrodynamical simulations of three different cosmological comoving volumes of size $(205 h^{-1}\ \text{cMpc})^{3}$ (TNG300),  $(75 h^{-1}\ \text{cMpc})^{3}$ (TNG100) and  $(35 h^{-1}\ \text{cMpc})^{3}$ (TNG50). The TNG simulations are run using the moving-mesh code \textsc{arepo} \citep{Springel2010} and employ prescriptions for various physical processes relevant for galaxy formation and evolution, such as star formation in dense regions, stellar evolution, gas heating by an evolving UV background, gas cooling through primordial and metal-line radiation, the seeding and growth through mergers and accretion of black holes, and feedback from galactic winds and AGN. All the details of the physics implemented in TNG can be found in \citet{TNGMethodsWeinberger2017,TNGMethodsPillepich2018}. All TNG runs assume a flat $\Lambda$CDM cosmology with cosmological parameters adopted from the \citet{Planck2015} results: $h=0.6774$, $\Omega_{\text{m}}=0.3089$, $\Omega_{\Lambda}=0.6911$, $\Omega_{\text{b}}=0.0486$, $n_{\text{s}}=0.9677$, and $\sigma_{8}=0.8159$. For each run, there are 100 outputs, approximately equally spaced in cosmic time from $z\sim20$ to $z=0$.

For this study, we focus on the data from the TNG50 simulation \citep{TNG50Nelson2019,TNG50Pillepich2019}, the highest-resolution run in the suite. The $(35 h^{-1}\ \text{cMpc})^{3} \approx (51.7\ \text{cMpc})^{3}$ volume initially contains $2160^{3}$ dark matter particles of mass $m_{\text{DM}}=4.5\times10^{5}\MSUN$ and an equal number of gas cells of mass $m_{\text{b}}=8.5\times10^{4}\MSUN$. Stellar particles are produced from the gas cells, inheriting similar masses, although these can subsequently decrease through stellar evolution and mass/metal return to the surrounding ISM gas. The stellar and DM particles have gravitational softening lengths of 288 pc at $z=0$, comoving for $z>1$ and physical at lower redshifts. The gas cells have adaptive softening lengths, set to 2.5 times the comoving radius of the cells, with a minimum value of 73.8 pc and an average value of $\sim70-150$ pc in the star-forming regions of galaxies \citep{TNG50Pillepich2019}.

Galaxies are identified in the simulation in a two-step process: Haloes are first detected using a friends-of-friends (FOF) algorithm \citep{Davis1985} with a linking length of $b=0.2$ times the mean inter-particle distance, only taking DM particles into consideration. Baryonic elements are assigned to the haloes that their nearest DM particles belong to. Galaxies are then detected within the FOF haloes using the code \textsc{subfind} \citep{Springel2001,Dolag2009}, which identifies gravitationally bound structures using all particles belonging to the halo. Finally, merger trees are generated by tracking the baryonic content of the galaxies through consecutive snapshots using the \textsc{sublink} algorithm, as detailed in \citep{RodriguezGomez2015}.

\begin{figure*}
    \includegraphics[width=\linewidth]{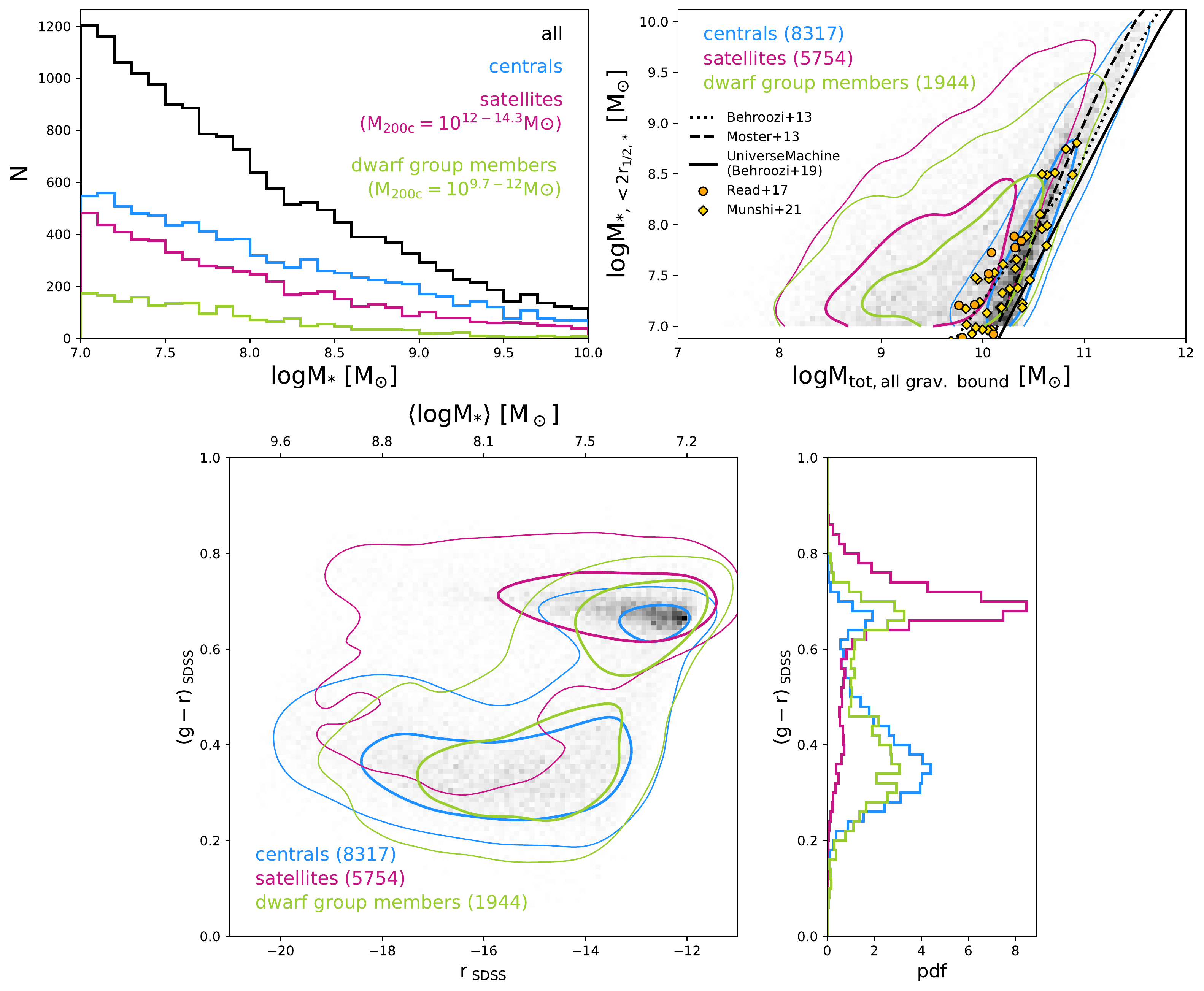}
    \caption{\textbf{TNG50 dwarf galaxy properties.} \emph{Top left}: Number distribution as a function of stellar masses for the full sample of TNG50 dwarf galaxies at $z=0$, and subdivided into central, satellite and dwarf-group subsamples. \emph{Top right}: Stellar mass-halo mass relation for the dwarfs showing the stellar mass $M_{\rm *,2r_{*,1/2}}$ vs. the total gravitationally bound mass of the galaxy $M_{\text{tot, all grav. bound}}$. The full sample is shown by the 2D histogram in grey; contours show the regions containing 50\% and 90\% of the three subsamples. For comparison, we show the best-fittting models from \citet{Moster2013} and \citet{Behroozi2013} as well as from the more recent UniverseMachine model \citep{Behroozi2019}. Additionally, orange circles show results from \citet{Read2017} for isolated dIrrs observed in the local Universe, while yellow diamonds show the results from \citet{Munshi2021} for simulated dwarfs, comprised of both field galaxies and satellites within MW-mass systems. \emph{Bottom}: Colour-magnitude diagram showing \emph{g-r} colour vs. r-magnitude (left) and the corresponding distribution of colours (right). The secondary \textit{x}-axis for the colour-magnitude diagram indicates the average log stellar mass in 2 mag. bins of $r_{\text{SDSS}}$. Note that the magnitudes used for these plots are those measured in \citet{TNGNelson2018}, using their Model C for dust attenuation.} \label{fig:sampleProperties}
\end{figure*}

\subsection{Extracting cumulative SFHs and stellar assembly times}
In order to extract the cumulative SFHs of a galaxy at $z=0$, i.e. here $\MSTAR(z)/\MSTAR(z=0)$, we calculate a weighted histogram of the stellar particles' birth ages in 50 Myr bins of cosmic time, where each particle is given a weight proportional to its \emph{present-day} mass. For the results presented in this paper, the cumulative SFHs are calculated using \emph{all gravitationally bound} stellar particles at any given time. 

Note that stellar particles can lose part of the mass they are initially formed with through mass return to the ISM such that their present-day mass is not equal to their birth mass. Furthermore, it should be noted that this method cannot account for any stellar mass that has been stripped away from galaxies prior to the time of inspection, which is possible for some of the satellite dwarfs. We also do not distinguish between in-situ and ex-situ star formation. The impact of all these methodology details are further discussed and quantified in an upcoming companion paper, but here we confirm that our results are nearly identical regardless of the choice of stellar particles mass. We have also verified that the results are qualitatively and quantitatively indistinguishable even if we consider only stellar particles within twice the 3D stellar half-mass radius, $2r_{\text{*,1/2}}$. This is quantitatively true for satellites of all stellar masses and centrals with $\MSTAR\gtrsim 10^{8.5}\MSUN$ galaxies and even for less massive centrals, the results remain qualitatively the same.

We further quantify the cumulative SFHs using three summary statistics, the times when each galaxy has built 10, 50 and 90 per cent of its total $z=0$ stellar mass by interpolating the cumulative SFH. These are referred to as the stellar assembly times $\tau_{10}$, $\tau_{50}$, $\tau_{90}$, respectively, and are in the form of lookback times, typically expressed in Gyrs.

\subsection{Measure of local environment and other galaxy properties}
There are several ways to characterize the environments of the dwarf galaxies e.g. the host mass of the FOF group they belong to, the radial position within the group in the case of satellites, the number density of galaxies around the dwarf or alternatively the distance to an $N^{\rm th}$ nearest neighbour. We choose to define the local environment of the dwarf galaxies by measuring the distance to the nearest massive host (in this study, chosen to have a stellar mass of $\MSTAR>10^{10}\MSUN$), although, in the case of satellites, we also explore the impact of the mass of the host clusters and the radial distance of the dwarf galaxies to their host centres in later sections. 

Unless specified otherwise, the stellar mass the dwarfs are identified with is always measured within twice the 3D stellar half-mass radius. Note that this is not the same mass as is used to normalize the SFHs (i.e. the mass of \emph{all} stellar particles). The difference in the two values is at most 0.16 dex for 90 per cent of the dwarfs in our sample and as high as 0.26 dex for some galaxies. The total mass of the underlying dark-matter (DM) FOF haloes or hosts is expressed in terms of $\MHOST$, the mass of all matter components enclosed within a spherical radius where the mean enclosed mass density is 200 times the critical density of the Universe.

%-------------------------------------------------

\section{The dwarf galaxy population of TNG50} \label{sec:TNG50dwarfs}

We select dwarf galaxies from the TNG50 $z=0$ snapshot within a stellar mass range of $\MSTAR=10^{7-10}\MSUN$, regardless of whether they are centrals or satellites within their FOF hosts. The lower mass limit ensures that the galaxies are resolved by $\gtrsim120$ stellar particles. Although the upper mass limit we adopt may be higher than more usual definitions of dwarf galaxies, it allows us to probe the transition in behaviour from dwarfs to more massive galaxies. The mass of $\MSTAR\sim10^{10}$ is also seen to be a critical transition mass for several key galaxy scaling relations, above and below which different physical mechanisms are seen to dominate the evolution of galaxies. 

Some subhaloes identified by \textsc{subfind} are found to be clumps of baryonic matter that are unlikely to be galaxies of cosmological origin, i.e. formed at the centre of their DM haloes. We use the `SubhaloFlag' defined in \citet{TNGDRNelson2019} to exclude such subhaloes from the sample. The corresponding criteria leave us with 16157 dwarfs, of which 8330 are the centrals of their FOF halo, while the remaining are satellites. Of these, we remove 142 dwarfs (13 of which are centrals), which could not be traced back to $z=2$. 

We further distinguish between satellites within hosts of mass $\MHOST>10^{12}\MSUN$ and $\MHOST<10^{12}\MSUN$ and refer to the latter as being part of a `dwarf group', whereas the former are referred to as `satellites' for the remainder of the paper. Note that the central dwarfs are found within FOF haloes of total mass $\MHOST=10^{8-12}$, so that there is significant overlap between the environments of the dwarf group and central galaxies. On the other hand, the satellites are found in much more massive hosts, by design. Within the TNG50 volume, we can sample two host clusters of total mass $\MHOST\gtrsim10^{14}\MSUN$ (comparable to Virgo), 6 (16) host groups with  $\MHOST\sim10^{13.5-14}\MSUN$ ($\MHOST=10^{13-13.5}\MSUN$) similar to Fornax (Cen-A), and 183 host groups with $\MHOST=10^{12-13}\MSUN$. The final dwarf sample therefore contains 8317 centrals, 5754 satellites and 1944 dwarf group members, overall spanning a great diversity of environments.

Fig.~\ref{fig:sampleProperties} shows some key properties of the TNG50 sample of dwarf galaxies. The top left panel shows the distribution of stellar masses of the full TNG50 sample, as well as the central, satellite and dwarf group subsamples. There does not appear to be any significant bias in stellar mass between the three subsamples, with the number of galaxies being approximately linearly increasing with decreasing log stellar mass. In the top right panel of Fig.~\ref{fig:sampleProperties}, we show the stellar mass-halo mass (SMHM) relation for the dwarfs, where the `halo mass' in this case is measured as the total mass of \emph{all} gravitationally bound particles belonging to a galaxy, and so for satellites, it represents their dynamical mass and not the virial mass of their hosts. The centrals have a relatively narrow SMHM relation, whereas both the satellite and dwarf group subsamples show significantly lower total masses at constant stellar mass. The satellites (and dwarf group satellites) are expected to have lost considerable amounts of dark matter after accretion onto their host groups and clusters while retaining a larger proportion, if not all, of their stellar mass, resulting in significantly higher stellar-to-halo mass ratio for satellites compared to centrals, as has been shown by several previous studies (e.g. \citealt{Niemiec2017,Sifon2018,Niemiec2019,Joshi2019,Engler2020,Dvornik2020}; however see \citealt{VanUitert2016} who find little difference between central and satellite SHMRs). For comparison with the centrals, we show the SMHM relations from \citet{Behroozi2013}, \citet{Moster2013} and the more recent UniverseMachine models \citep{Behroozi2019}, using their best-fitting models to observational data (in the case of the UniverseMachine model, we use their parameters for observations of all centrals). Additionally, we show results from \citet{Read2017} for isolated dIrrs which they obtain by fitting HI rotation curves for the galaxies (orange circles), as well as from \citet{Munshi2021} for simulated dwarfs both in the field and around MW-mass galaxies (yellow diamonds). Note that these are approximate comparisons, as we have not measured stellar masses or dynamical masses in exactly the same manner as either of these authors. Nonetheless, we have confirmed that considering $\MHOST$ instead of $M_{\rm tot,\ all\ grav. bound}$ or the total bound stellar mass rather that only within the 3D stellar half-mass radius has a small impact on the SMHM relation ($\Delta\MSTAR\sim0.002-0.1$ dex at fixed halo mass on average in this mass regime, with the choice of stellar mass having the largest impact).

In the bottom panels of Fig.~\ref{fig:sampleProperties}, we show the colour-magnitude diagram in the SDSS \textit{g} and \textit{r} bands (left), as well as the distribution of \textit{g-r} colour (right) for the dwarf galaxies. Galaxy photometry is taken from the measurements by \citet{TNGNelson2018}, using their Model C to account for dust attenuation, and considering particles within 30~pkpc. The figure shows that the satellites in massive hosts are predominantly located on the red sequence region of the diagram, while centrals occupy the blue cloud region, as expected. The dwarf group members appear to be an intermediate population, occupying both these regions on the colour-magnitude diagram. Similarly, we have confirmed (not shown) that while the majority of the centrals lie on a star-forming main sequence, the majority of the satellites are quenched, in line with previous results based on TNG. The reader can refer to \citet{Donnari2020a, Donnari2020b} for an extensive quantification of the quenched fractions of galaxies as a function of galaxy stellar mass, host mass, halo-centric distance and time since infall, and of the level of agreement between TNG simulations and observations. In the case of the dwarf group subsample, although the majority are star-forming, there is a significant portion of the sample that is quenched, especially at the low-mass end.

%-------------------------------------------------

\section{Cumulative SFHs from TNG50} \label{sec:SFHs}

\subsection{Centrals vs. satellites}

\begin{figure*}
    \centering
	\includegraphics[width=0.5\linewidth]{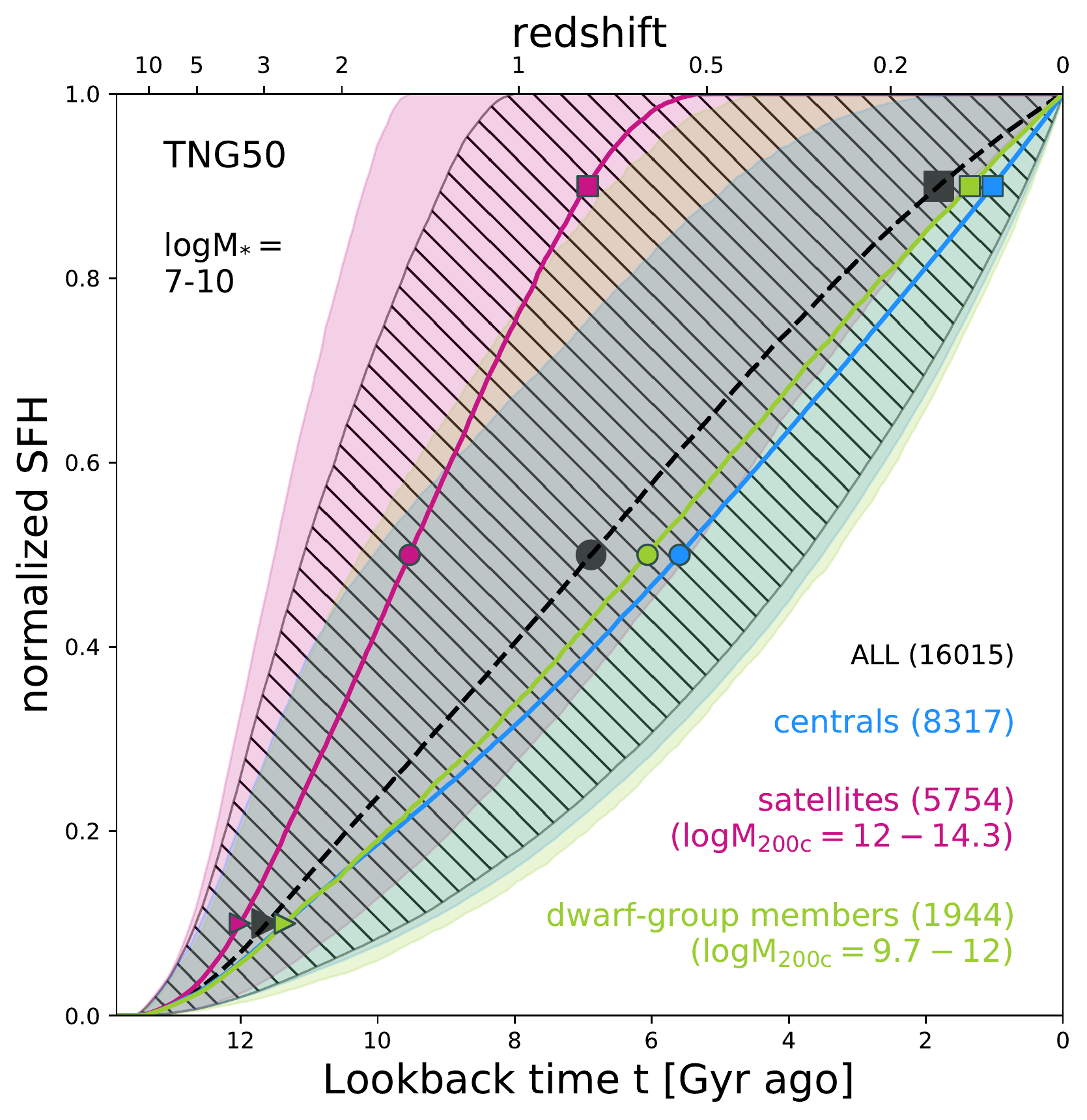}
	\includegraphics[width=0.9\linewidth]{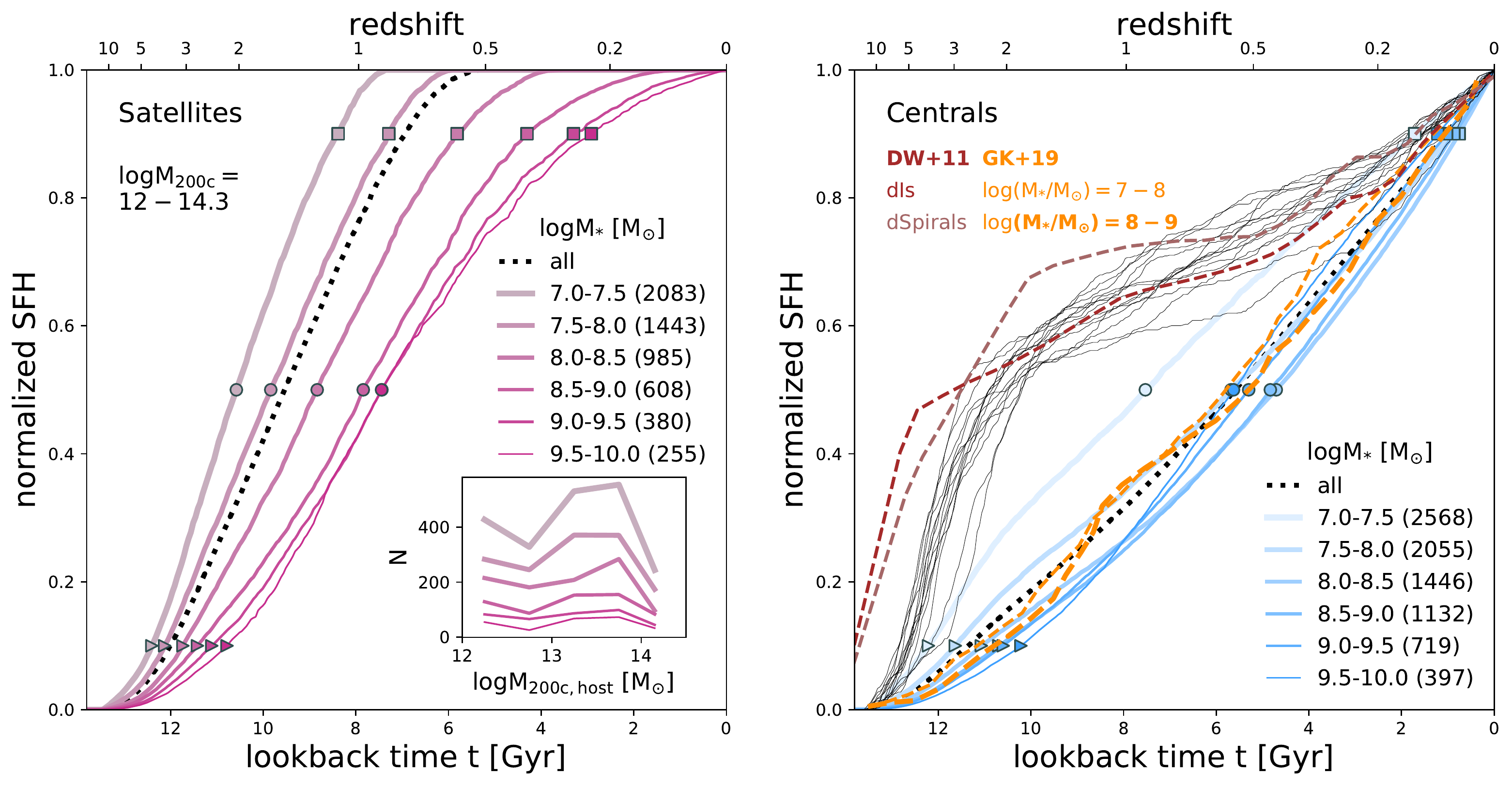}
    \caption{\textbf{Global cumulative SFHs and mass dependence of TNG50 dwarf galaxies at $z=0$.} \emph{Top}: Cumulative SFHs for the full dwarf sample (black dashed curve and hatching) and for the central, satellite and dwarf group subsample (coloured curves and shaded areas). The curves show median values and the hatching and shaded areas show the 10$^{\rm th}$-90$^{\rm th}$ percentiles. \emph{Bottom}: Cumulative SFHs for the satellite and central subsamples, in bins of stellar mass. The dashed curve in each panel shows the results for the full subsample for reference. In all panels, triangle, circle and square markers indicate when the subsample has assembled 10\%, 50\% and 90\% of the total $z=0$ stellar mass respectively. The inset in the bottom left panel shows the distribution of the satellites' host masses in the same stellar mass bins, reflecting the volume-limited nature of the TNG50 dwarf sample. Additionally, we show simulation results GK19 and observational results from DW11 respectively for comparison; the thin (thick) dashed orange curve represents central dwarfs in the mass range of $\MSTAR=10^{7-8}\MSUN$ ($10^{8-9}\MSUN$) from GK19. Brown dashed curves show the average results from DW11 for dIrrs and dSpirals (ee text for details). Thin black curves in the bottom right panel show the cumulative SFHs for 15 centrals, selected within the mass range of $\MSTAR=10^{7.5-8.5}\MSUN$, and with $\tau_{50}>10$ Gyr ago and $\tau_{90}<2$ Gyr ago, to highlight a few analogues the of DW11 results.} \label{fig:cumulSFHsAndMstarBins}
\end{figure*}

In the top panel of Fig.~\ref{fig:cumulSFHsAndMstarBins}, we first compare the stacked cumulative SFHs of the entire TNG50 dwarf sample and the central, satellite and dwarf group subsamples. In order to combine the SFHs of multiple galaxies, we show the median of individual cumulative SFHs in each bin of lookback time. Hatching and shaded areas denote the $10^{\rm th}-90^{\rm th}$ percentiles, and thus quantify the galaxy-to-galaxy variation or scatter. In all panels, triangle, circle and square markers indicate when the subsets have assembled 10, 50, and 90 per cent of the total $z=0$ stellar mass, i.e. the average stellar assembly times $\tau_{10}$, $\tau_{50}$, $\tau_{90}$, respectively.

The figure shows that satellites of massive hosts ($\MHOST=10^{12-14.3}\MSUN$) built up 90 per cent of their present-day stellar mass approximately 7~Gyr ago i.e. at a redshift of $z\sim0.75$, whereas the central dwarfs did so around 1~Gyr ago at $z\sim0.075$. Additionally, the scatter in the cumulative SFHs is significantly greater for the satellite dwarfs, and the majority of them assembled \emph{all} their stellar mass over 6~Gyr ago, i.e. at $z\gtrsim0.5$. The dwarf group members show a similar median SFH to the centrals, although they have a slightly larger scatter, such that a significant portion of the dwarf group galaxies built up all their stellar mass over 4~Gyr ago ($z\gtrsim0.35$), consistent with a noticeable fraction of them being red and quenched at $z=0$. Since the central and dwarf group subsamples account for about 64 per cent of the total sample, the median cumulative SFH for the full sample appears similar to that of the centrals and the dwarf group galaxies.

In the lower panels of Fig.~\ref{fig:cumulSFHsAndMstarBins}, we further investigate the dependence of the cumulative SFHs on stellar mass, separately for the satellite and central subsamples. For the remainder of the paper, we do not show the results for the dwarf group subsample separately, since their average results are nearly identical to those of the central dwarfs. In the case of the satellites of TNG50 massive hosts, the cumulative SFHs are strongly dependent on the stellar mass of the galaxies, with more massive galaxies building their stellar mass at later times -- 90 per cent of the stellar mass is assembled by $\sim3$~Gyr ago for the most massive bin of galaxies, by $\gtrsim8$~Gyr ago for the least massive. In contrast, only the least massive centrals show any significant deviation in their cumulative SFHs compared to the remaining centrals, with the largest difference being the time at which they had assembled 50 per cent of their stellar mass. However, as we show in what follows and in Appendix~\ref{sec:appResEffects}, such a departure of the lowest mass central galaxies in TNG50 ($\lesssim10^{7.5}\MSUN$) from the typical cumulative SFHs of centrals is mostly due to resolution effects. Despite this and crucially, it is clear from Fig.~\ref{fig:cumulSFHsAndMstarBins} that, to zeroth order, the cumulative SFHs are determined by whether the dwarfs are satellites in hosts of considerable mass ($>10^{12}\MSUN$) or centrals, followed by stellar mass in the case of the satellites. 

It is important to note that the median curves in Fig.~\ref{fig:cumulSFHsAndMstarBins} are the results for a volume-limited sample of hosts and subhaloes as they naturally form in a $\Lambda$CDM model and as they are populated by galaxies that follow the effective SMHM relation emerging from the TNG galaxy formation model. The inset in the lower left panel of Fig.~\ref{fig:cumulSFHsAndMstarBins} shows the number of TNG50 dwarf satellites as a function of host mass; simulated and observed galaxy samples with different host and satellite stellar mass distributions can, in principle, return different average cumulative SFHs. Keeping this in mind, in the lower right panels, we show analogous results from \citet{Weisz2011} and \citet{GarrisonKimmel2019} (DW11 and GK19), from observations and simulations, respectively (similar comparisons for the satellites are shown in Section~\ref{sec:environmentalFactors}). The dashed brown curves show the average SFHs for field dwarfs of two morphological types from DW11 (fig. 6 in the paper): dIs and dSpirals. The average stellar masses for each of these subsamples are $7.4\times10^{7}\MSUN$ and $7.5\times10^{8}\MSUN$ respectively. We chose to show these two types since the majority of them are likely to be centrals, based on the measurements of the galaxies' isolation by DW11. The thin (thick) dashed orange curve represents the average SFH for central dwarfs from GK19 (fig. in the paper) of mass $\MSTAR=10^{7-8}\MSUN$ ($\MSTAR=10^{8-9}\MSUN$). 

A note of caution here is that these previous results should not be compared directly to the TNG50 results, especially in the case of observations, as we have not attempted to produce any mock/synthetic versions of them. In the case of simulations, although the methods may be comparable, it should be kept in mind that there may still be small differences in the choice of galaxy extent and in the measurement of galaxy masses, for example. Nevertheless, these comparisons may be useful to give context to our simulation results. While we find overall good agreement between our results and the simulation results of GK19, the results of DW11 are significantly different from the TNG50 averages, with their observed dwarfs having a period of much higher SF activity at earlier epochs in cosmic evolution compared to the TNG50 results. However, contrary to what may be the impression from these average results, we do find several galaxies in TNG50 whose SFHs are individually similar to those observed by DW11, as shown in the lower right panel of Fig.~\ref{fig:cumulSFHsAndMstarBins}. Thin black curves show 15 TNG50 central dwarfs with $\MSTAR=10^{7.5-8.5}\MSUN$, and with $\tau_{50}>10$ Gyr ago and $\tau_{90}<2$ Gyr ago, to highlight a few analogues of the results of DW11. In fact, the sample sizes for DW11 (as well as GK19) are significantly smaller than those from TNG50, and therefore, these discrepancies may at least partially be the result of increased scatter due to low-number statistics. As is evident, even within TNG50 there are some galaxies whose individual SFHs are very different from the median curves at the same stellar mass and are possibly more similar to the ones observed in the local Universe. It remains to be determined whether an actual inconsistency is in place between the observations within the Local Volume and the model underlying TNG50; this would require a careful statistical analysis and mocking both the observations and the selection of the observed galaxies, a task that is beyond the scope of this paper but which will be tackled in the future.

\begin{figure*}
    \centering
    \includegraphics[width=\linewidth]{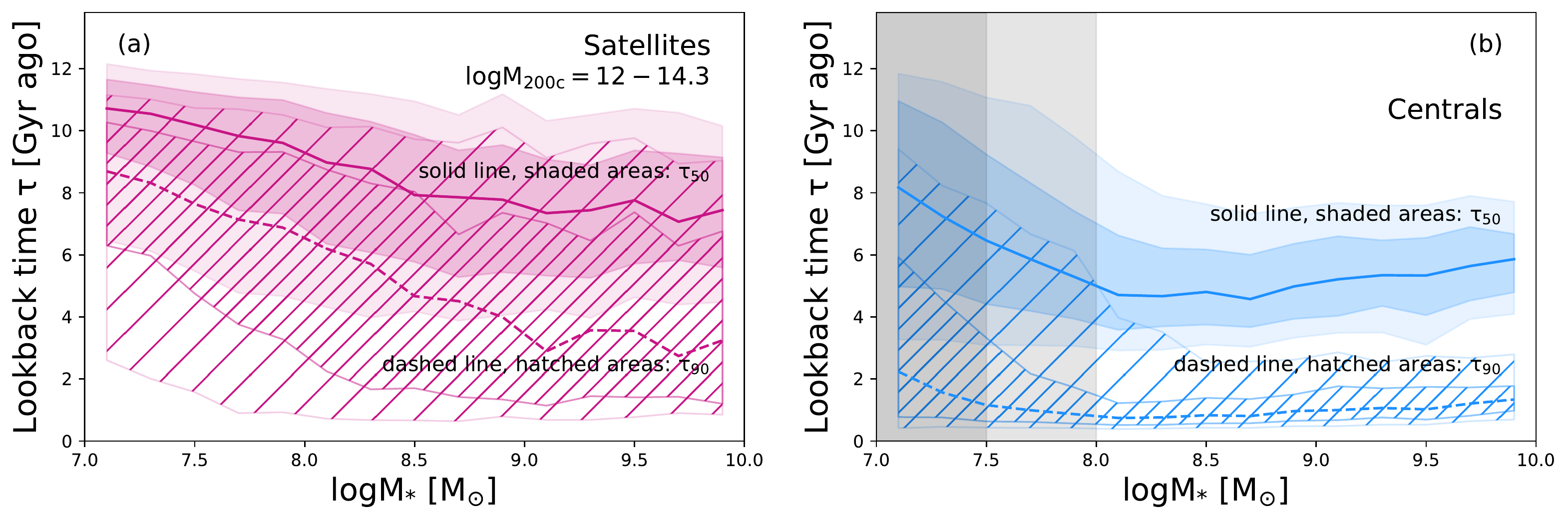}
    \setlength{\tabcolsep}{0pt}
    \begin{tabular}{c|c}
    \includegraphics[width=0.5\linewidth]{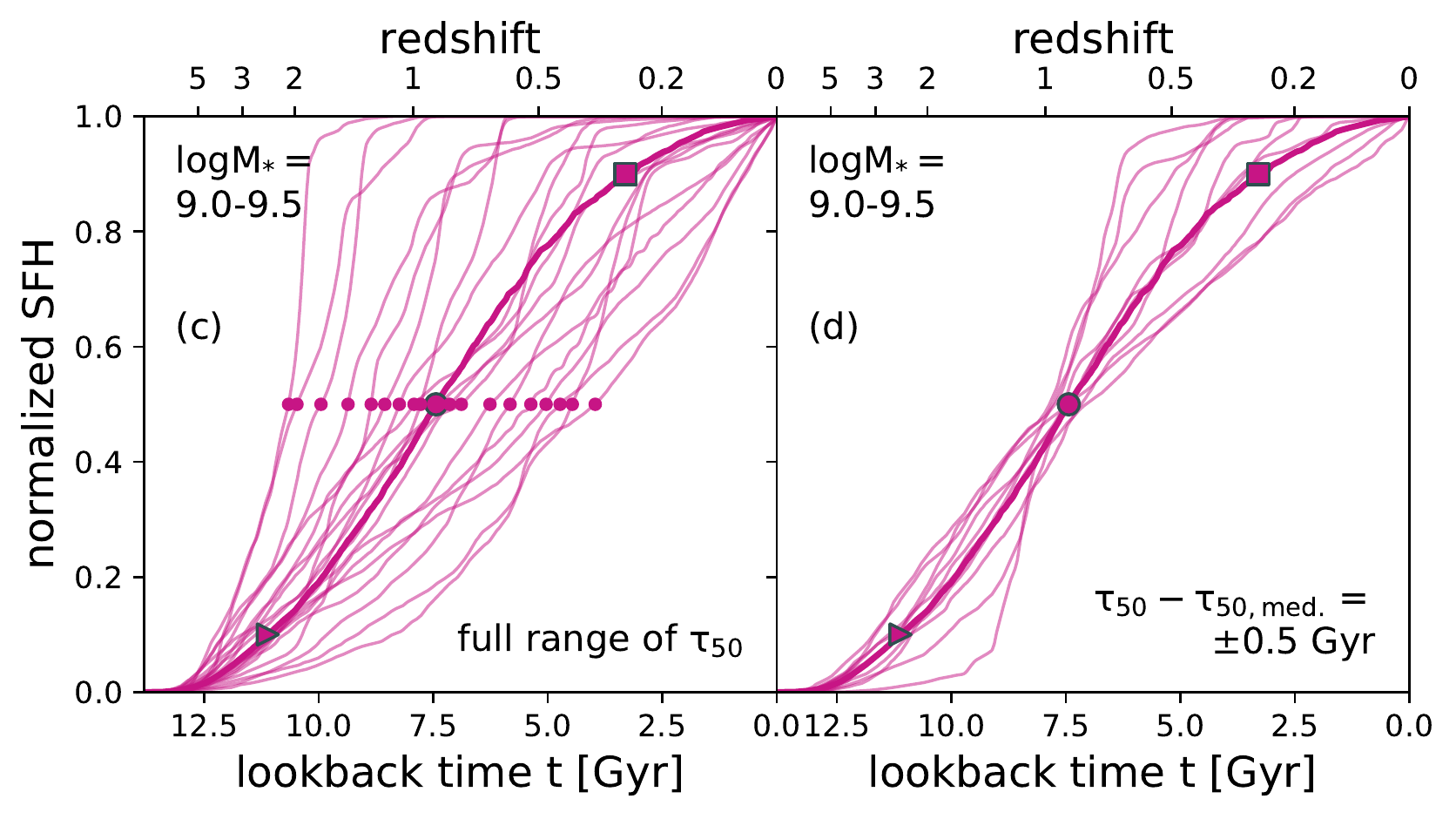}
    \includegraphics[width=0.5\linewidth]{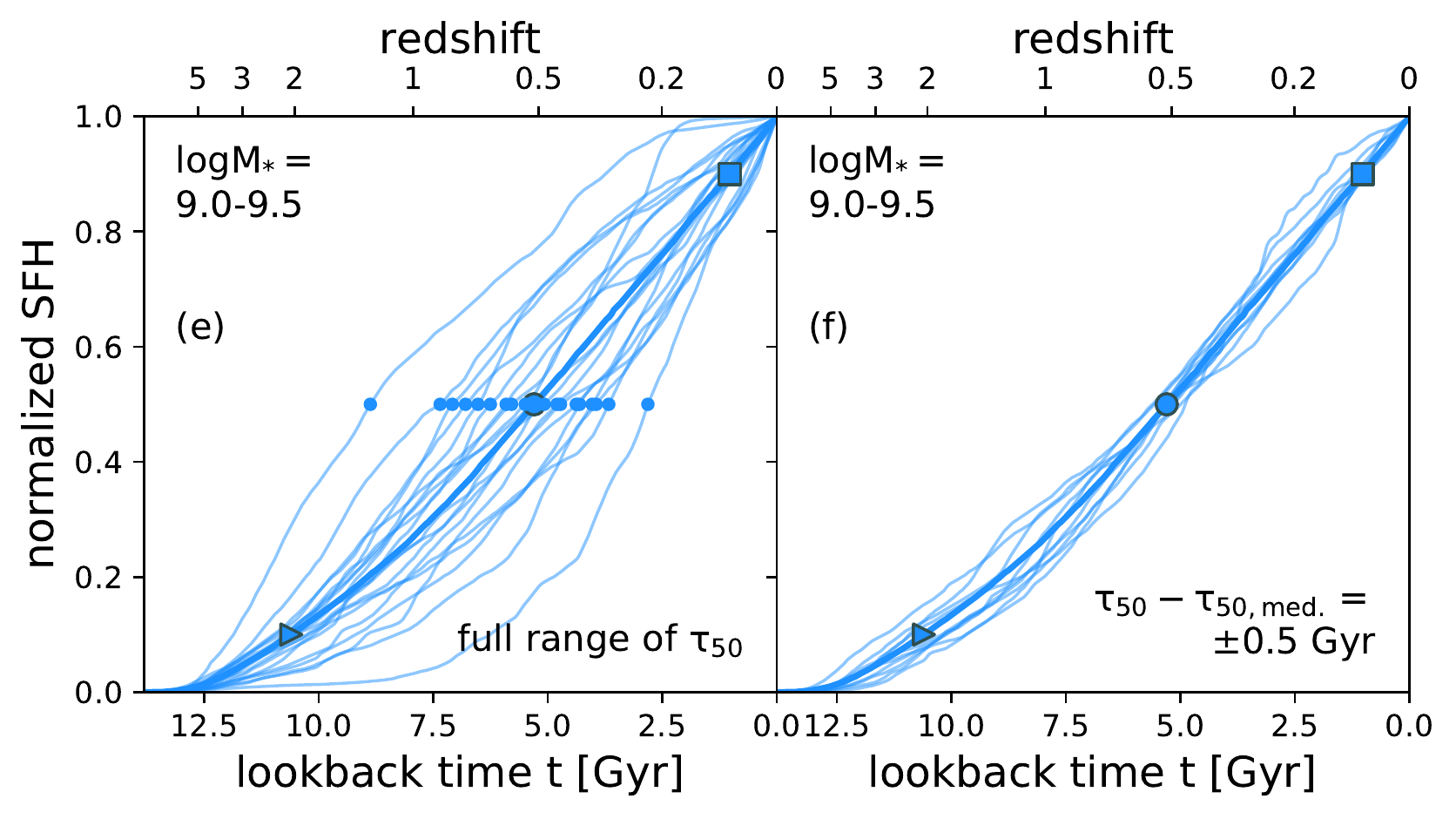}
    \end{tabular}
    \caption{\textbf{\emph{(Top)}: Distribution of stellar assembly times of TNG50 dwarfs.} Median values of $\tau_{50}$ (solid curves) and $\tau_{90}$ (dashed curves) as a function of stellar mass for satellites (left) and centrals (right). The darker shaded and hatched areas show the corresponding $20^{\rm th}-80^{\rm th}$ percentile scatter; the lighter areas show the $5^{\rm th}-95^{\rm th}$ percentile scatter. The darker (lighter) gray area in panel (b) indicates where the masses where the results are not converged to better than 2 (1) Gyr; at higher masses the results are well converged. \textbf{\emph{(Bottom)}: Diversity of SFHs.} In each panel, thick curves represent the median cumulative SFH for all dwarfs of the given subsample within the mass range of $\MSTAR=10^{9-9.5}\MSUN$; thin curves show individual cumulative SFHs. \emph{Panels (c) \& (e)}: Individual SFHs for satellite (panel c) and central (panel d) dwarfs, chosen randomly within 5 percentile bins of $\tau_{50}$. \emph{Panels (d) \& (f)}: Individual SFHs for 10 randomly chosen dwarfs but restricted to have $\tau_{50}=\tau_{50,\mathrm{med.}\pm0.5}$~Gyr, where $\tau_{50,\mathrm{med.}}$ is the $\tau_{50}$ for the median SFH.} \label{fig:t50t90AssemblyTimes}
\end{figure*}

\subsection{Galaxy-to-galaxy diversity}

While Fig.~\ref{fig:cumulSFHsAndMstarBins} shows the median cumulative SFHs as a function of stellar mass, there can be significant scatter in the individual SFHs. In panels (a) and (b) of Fig.~\ref{fig:t50t90AssemblyTimes}, we present $\tau_{50}$ and $\tau_{90}$ as a function of galaxy stellar mass, to show not only the typical values, but also to examine the scatter in both quantities. There is a clear correlation between $\tau_{90}$ and stellar mass for the satellites, with less massive satellites assembling their stellar mass at earlier times, whereas for the centrals, only a weak trend is seen at the lowest masses; for $\MSTAR\gtrsim10^{8}\MSUN$, $\tau_{90}$ ($\tau_{50}$) shows a mild increase with stellar mass from $\sim0.7-1.3$~Gyr ago ($\sim4.7-5.9$~Gyr ago). The trends in $\tau_{50}$ are similar, but somewhat weaker for the satellites. In fact, the upturn of the stellar assembly times at the low-mass end for the centrals is an artifact of the finite numerical resolution, so that, at the resolution of TNG50, $\tau_{90}$ ($\tau_{50}$) is converged with an error of 1~Gyr (2-2.5~Gyrs) for galaxies below about $10^{7.5}\MSUN$ -- see Appendix~\ref{sec:appResEffects} for further details and discussion. Hence, shaded gray areas in the top right panel of Fig.~\ref{fig:cumulSFHsAndMstarBins} denote regimes where the results from TNG50 central dwarfs should be interpreted with caution. Note that the results for satellites in the TNG model are resolution independent across the satellite and host mass ranges studied here (see Appendix~\ref{sec:appResEffects}).

In the case of the satellites, there is significant scatter in the values of both $\tau_{50}$ and $\tau_{90}$ that mildly increases with stellar masses, with $\bar{\sigma}_{\tau_{50}}\sim2.0-3.0$~Gyr and $\bar{\sigma}_{\tau_{90}}\sim3.3-4.2$~Gyr, where $\sigma_{\tau}$ is defined as half the difference between the $10^{\rm th}$ and $90^{\rm th}$ percentile values of the given critical time in stellar mass bins of 0.25 dex. In contrast, the scatter in $\tau_{50}$ and $\tau_{90}$ is significantly lower for centrals with $\MSTAR>10^{8}\MSUN$ compared to that of satellites in the same mass range -- approximately constant with stellar mass at values of $\bar{\sigma}_{\tau_{50}}\sim1.8$~Gyr and $\bar{\sigma}_{\tau_{90}}\sim0.75$~Gyr. At masses of $\MSTAR=10^{7-8}\MSUN$ however, there is significantly more scatter in the $\tau_{50}$ and $\tau_{90}$ values, as high as $\sim3.8$~Gyr and $\sim3.6$~Gyr respectively. Overall, the average scatter in $\tau_{90}$ values for the centrals is higher at lower resolution, with the increase in scatter at lower masses seen at all resolutions. In the case of the $\tau_{50}$ values, there is no obvious monotonic trend between scatter and resolution and in the case of the centrals, the degree of scatter is nearly identical at all resolutions.

It should be kept in mind that the vast majority of centrals are star forming at $z=0$, and therefore, it is expected that they built up 90 per cent of their stellar mass quite recently. In contrast, the vast majority of satellites are quenched and therefore, will have reached $\tau_{90}$ at earlier times. The larger scatter in their values of $\tau_{90}$ are likely a result of different accretion times, orbital parameters and quenching times, as we show in Section~\ref{sec:environmentalFactors} and discuss in Section~\ref{sec:discQuenching}.

In the bottom panels of Fig.~\ref{fig:t50t90AssemblyTimes}, we further examine the diversity of the cumulative SFHs and how representative the median SFHs are of individual SFHs. In panels (c) and (e), we reproduce the median cumulative SFHs for the satellites and centrals respectively, with $\MSTAR=10^{9-9.5}\MSUN$ from Fig.~\ref{fig:cumulSFHsAndMstarBins}. Additionally, with thin curves, we show the individual SFHs of 20 dwarfs, each chosen randomly from within 5 percentile bins of $\tau_{50}$ for the given mass range. The $\sim6$~Gyr scatter in $\tau_{50}$ for the satellites in this mass range corresponds to a wide variety of SFHs, from those that were quenched over 7~Gyr ago to those that are still star forming. Similarly, the $\sim4$~Gyr scatter for the corresponding centrals is due to a range of possible SFHs; although most of these centrals are star forming, individual galaxies can build up their stellar mass at vastly different rates and epochs. 

In panels (d) and (f) of Fig.~\ref{fig:t50t90AssemblyTimes}, we show the same median SFHs as panels (c) and (e); additionally, thin curves show the individual SFHs for 10 satellite and central dwarfs, respectively, chosen randomly but restricted to $\Delta\tau_{50}=\tau_{50}-\tau_{50,\mathrm{med.}}=\pm0.5$~Gyr, where $\tau_{50,\mathrm{med.}}$ is the $\tau_{50}$ value for the median SFH for the subsample. Approximately 21 (28) per cent of the satellites (centrals) in this mass range fall within this range of $\Delta\tau_{50}$. In the case of the centrals, the median SFH is similar to the individual SFHs with similar $\tau_{50}$ values. In contrast, for the satellites, even within this narrow range of $\tau_{50}$ values, the individual SFHs are varied both at early and late times. Different accretion and orbital histories within their host environments result in significantly different SFHs in recent epochs. The variation in earlier epochs also suggests a range of SFRs at early times.

These results show that the median cumulative SFHs are representative of the TNG50 galaxy populations only in an average sense and individual, randomly selected galaxies can exhibit vastly different SFHs. In the following sections, we explore various drivers of such a diversity in SFHs, by focusing on the impact of environment. The role of intrinsic galaxy properties and their correlations with the cumulative SFHs of galaxies is investigated in an upcoming, companion paper. 

%-------------------------------------------------

\section{The role of environment} \label{sec:environmentalFactors}

\begin{figure*}
    \includegraphics[width=\linewidth]{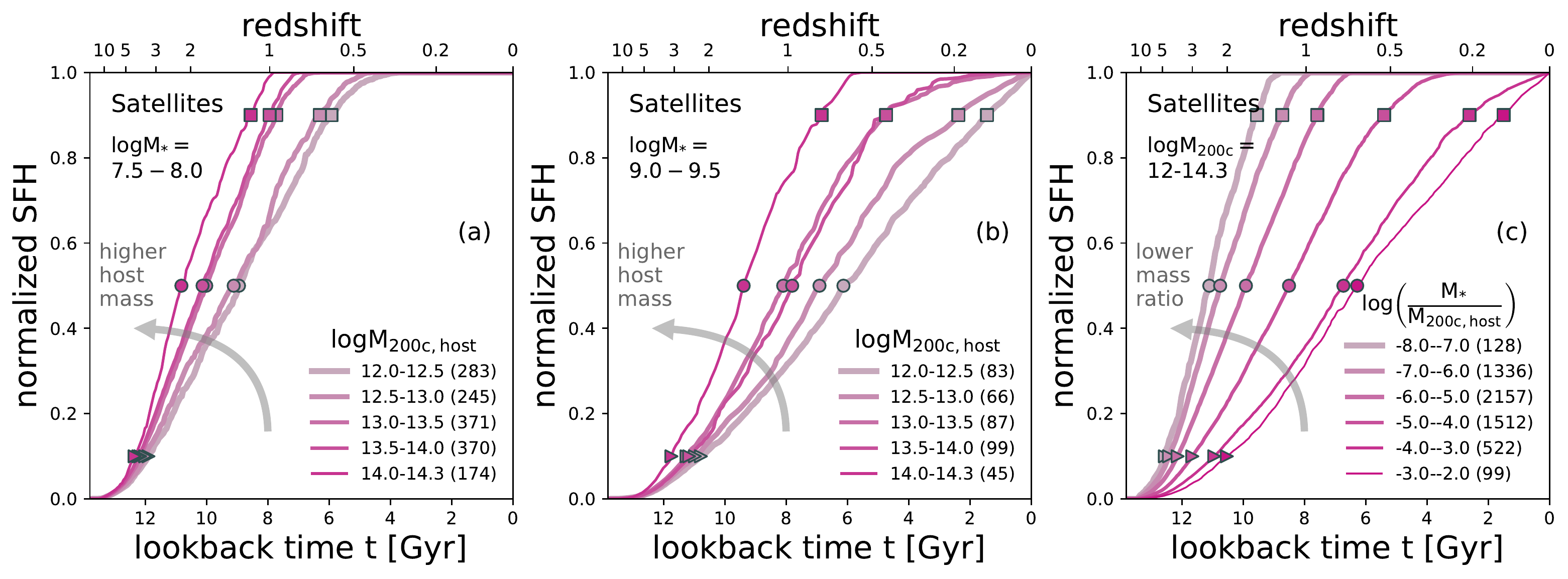}
    \includegraphics[width=\linewidth]{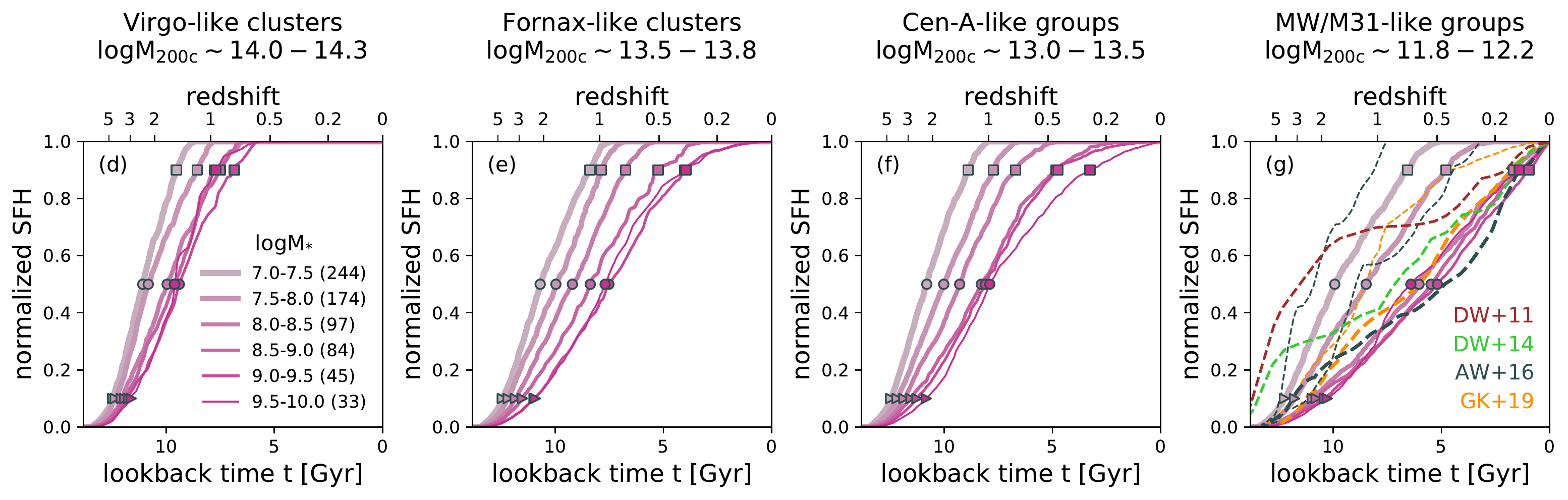}
    \caption{\textbf{Dependence on host mass of TNG50 dwarfs at $z=0$.} \emph{Panels (a) \& (b)}: Cumulative SFHs for satellites in bins of host mass $\MHOST$, for low-mass ($\MSTAR=10^{7.5-8}\MSUN$) and high-mass ($\MSTAR=10^{9-9.5}\MSUN$) galaxies. \emph{Panel (c)}: Cumulative SFHs for satellites in bins of stellar mass to host mass ratio $\MSTAR/\MHOST$. \emph{Panels (d-g)}: Cumulative SFHs for satellites in various environments similar to groups and clusters in the nearby Universe, further subdivided in bins of stellar mass. In the panel (g), we show comparable results from AW16 (gray), GK19 (orange) which are from hydrodynamical zoom simulations, and DW11 (gray) and DW14 (teal), which are observational. For AW16 and GK19, thinner curves are for dwarfs with $\MSTAR=10^{7-8}\MSUN$, thicker curves for $\MSTAR=10^{8-9}\MSUN$. The AW16 results shown here are cumulative SFHs for individual dwarfs. From DW11, we show their results for dSphs. See text for details.} \label{fig:satHostMass}
\end{figure*}

As seen in the previous section, the SFHs of the dwarfs are significantly dependent on whether they are satellites or centrals and on their stellar mass. However, other environmental factors can also play a significant role in determining the SFHs of these galaxies, especially in the case of satellites. In the following sections, we explore the effect of a few environment indicators -- for satellites, the mass of the host halo they belong to, their clustercentric distance and their accretion time, and for both centrals and satellites, the distance to the nearest massive neighbouring galaxy.

\subsection{Satellite dwarfs: impact of host mass, position and accretion time}

\subsubsection{Host mass}

In Fig.~\ref{fig:satHostMass} (a) and (b), we show the cumulative SFHs for the satellite dwarfs in bins of $z=0$ host mass $\MHOST$, for dwarfs with $\MSTAR=10^{7.5-8}\MSUN$ and $\MSTAR=10^{9-9.5}\MSUN$, to show the effects on low and high mass dwarfs separately. We choose to show these bins instead of the lowest and highest stellar mass bins in our sample to circumvent any possible resolution effects at the low mass end (see however Appendix~\ref{sec:appResEffects}) and to have a sufficiently large sample at the high mass end to show robust trends at the high mass end. In addition to stellar mass, the SFHs of the satellites at fixed stellar mass show a clear dependence on host mass: satellites in more massive hosts have built up their stellar mass at earlier times, or, in other words, satellites in more massive hosts may have stopped forming stars at earlier times. Note that here, group/cluster membership is determined based on which FOF group the dwarf belongs to at $z=0$, with no constraints on its distance to the group/cluster centre and with no distinction between pre-processed or direct-infaller satellites (see \citealt{Donnari2020a} for an extensive discussion). In fact, we have confirmed that for the most massive hosts with $\MHOST=10^{14-14.3}\MSUN$, the cumulative SFHs are similar regardless of stellar mass, with the median $\tau_{90}\sim7.5-9$~Gyr ago, whereas in the least massive hosts with $\MHOST=10^{12-12.5}$, $\tau_{90}$ can range from $\sim1.5-7$~Gyr ago across our $10^{7-10}\MSUN$ stellar mass range.

This competition between stellar mass and host mass is shown more explicitly in Fig.~\ref{fig:satHostMass} (d-g), where we give the TNG50 predictions for the median cumulative SFHs for satellites in hosts that are analogous in total mass to groups and clusters in the local Universe, further subdivided in bins of stellar mass. The figures show that in the most massive hosts in TNG50, i.e. Virgo-like clusters, on average, satellites of all masses built up the entirety of their current stellar mass over 7~Gyr ago, with only a weak dependence on stellar mass and no ongoing star formation on average. In the more intermediate Fornax-like and CenA-like hosts, more massive satellites built up their stellar mass at later times. Finally, in the lowest mass MW/M31-like systems, only the least massive satellites ($\MSTAR<10^{8}\MSUN$) have $\tau_{90}\sim6-8$~Gyr ago, whereas at higher masses, all satellites built up their total stellar mass only in the last couple of Gyrs. 

The dependence of the satellite SFHs on both stellar mass and host mass is well captured by their correlation with the ratio between satellite stellar mass and host total mass, as shown in Fig.~\ref{fig:satHostMass} (c), with lower mass-ratio satellites exhibiting earlier cumulative SFHs. We have verified (but do not show) that, at fixed mass ratio, residual dependencies of the cumulative SFHs on host mass or satellite stellar mass are negligible, so that the mass ratio is a good estimator of the cumulative SFHs of satellite dwarfs.

For the MW/M31-like systems, in Fig.~\ref{fig:satHostMass} (g), we also show comparable results from AW16 (gray) and GK19 (orange), both of which are from hydrodynamical zoom simulations, and DW11 (brown) and DW14 (green), which are observational. From AW16 (fig. 5 in their paper), we show the individual SFHs for the three most massive satellite dwarfs around a MW-mass host (within 300 kpc from the host) in their sample with stellar masses of $2\times10^{7}\MSUN$, $4\times10^{7}\MSUN$ (thinner curves) and $2\times10^{8}\MSUN$ (thicker curve). From GK19 (fig. 3 in their paper), we show the stacked cumulative SFH for their samples of satellites found within 300 kpc from an isolated MW-like host in the mass ranges of $\MSTAR=10^{7-8}\MSUN$ and $\MSTAR=10^{8-9}$. As in the lower right panel of Fig. \ref{fig:cumulSFHsAndMstarBins}, from DW11 (fig. 6 in their paper) we show the average cumulative SFHs of a single morphological type, dSphs, which have an average $\MSTAR=5.7\times10^{7}\MSUN$ and are likely to be satellites based on their isolation measurement. Finally, from DW14 (fig. 10 in their paper), we show the stacked cumulative SFH for their sample of LG dwarfs with $\MSTAR\sim10^{7-8}$. With the caveat that these comparisons are at face value (i.e. without rigorous matching of measurement methods), we find that the TNG50 results are in good qualitative agreement with the two simulation results, showing the trend of later stellar assembly for more massive dwarfs. In particular, the results from TNG50 are broadly in agreement with the results from GK19 in both mass bins of $\MSTAR=10^{7-8}\MSUN$ and $\MSTAR=10^{8-9}\MSUN$. The results from AW16 are individual SFHs for three of the 13 dwarfs in their sample whose stellar masses overlap with our sample. The low mass AW16 dwarf SFHs bracket the average SFH for our sample with $\MSTAR=10^{7-8}\MSUN$, but are within the scatter of our samples. On the other hand, the observational results appear to show later assembly times than for our TNG50 sample of dwarfs with $\MSTAR=10^{7-8}$. Note that we have confirmed that these findings are qualitatively the same whether or not we restrict the TNG50 results to only those satellites that reside within the virial radius of their hosts (as opposed to being part of a FOF host, as adopted throughout). Yet, it should be kept in mind that within the TNG50 volume we also find a large diversity of cumulative SFHs for the satellites, and the results of DW11 are consistent with TNG50's within this scatter. We postpone to a future paper the task of verifying the physical and statistical consistency between our and the observed results, e.g. by checking whether, within TNG50, we can find a MW or Andromeda analogue whose satellite population exhibit SFHs consistent with those measured for the Local Group by DW14. 

Thus, from Fig.~\ref{fig:satHostMass}, the key factor in determining the SFHs of satellite dwarfs appears to be stellar mass, except in the case of the most massive hosts, where the effect of the environment dominates. The large scatter seen in the cumulative SFHs for satellites of Fig.~\ref{fig:cumulSFHsAndMstarBins} is partly due to the large range of host masses the dwarfs belong to. A further contributor to this scatter is likely the diversity in orbits and accretion times of the satellites and their cluster-centric distance. We explore these factors in the following sections.

\begin{figure*}
    \includegraphics[width=\linewidth]{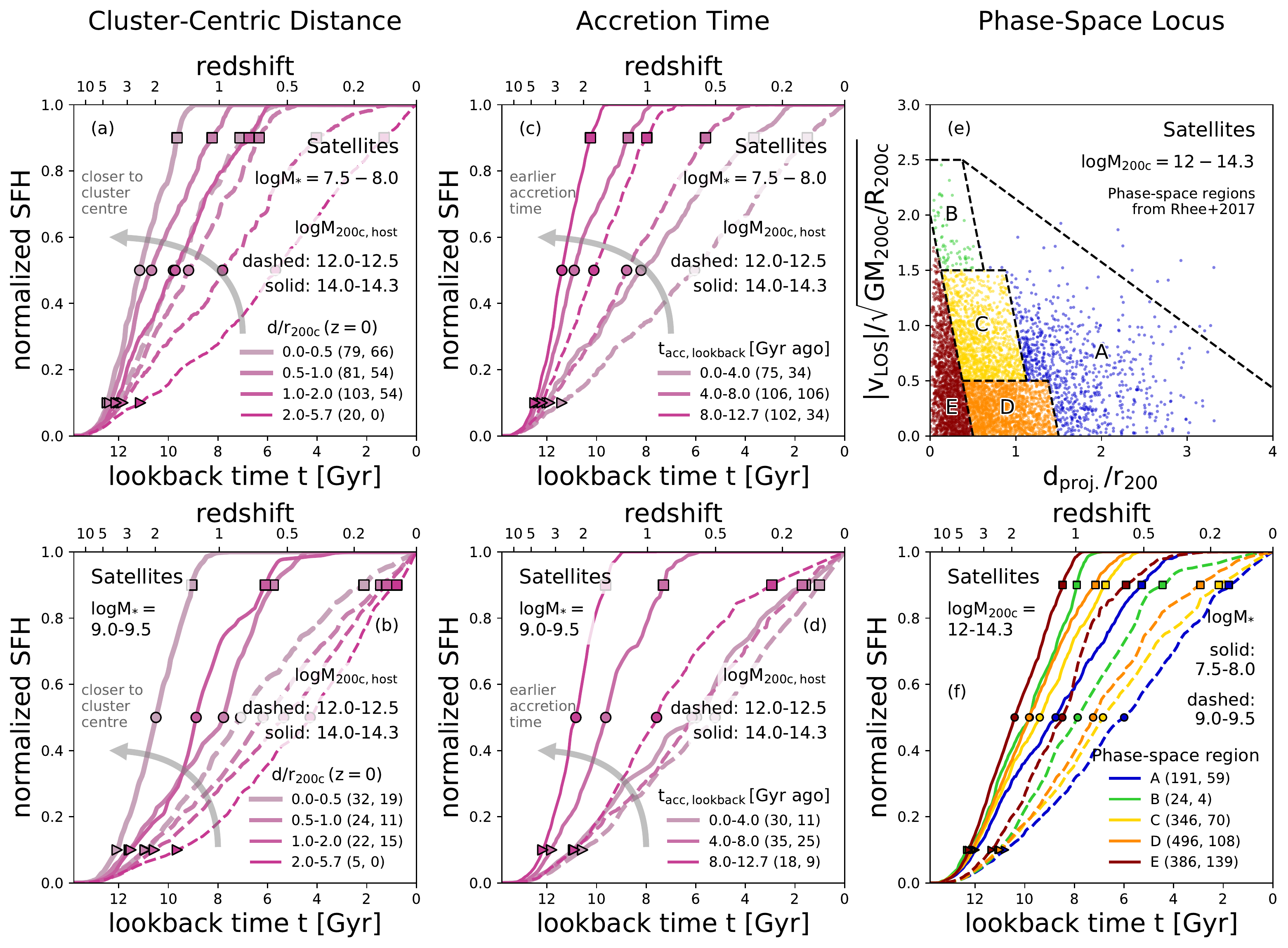}
    \caption{\textbf{Dependence on radial position, accretion time and phase-space locus of TNG50 dwarfs.} Cumulative SFHs for satellites in bins of $z=0$ cluster-centric distance (\emph{panels (a) \& (b)}) and accretion time (\emph{panels (c) \& (d)}) for the low-mass ($\MSTAR=10^{7.5-8}\MSUN$) and high-mass ($\MSTAR=10^{9-9.5}\MSUN$) dwarfs, further restricted to the least (dashed curves) and most (solid curves) massive bins of host mass. The cluster-centric distance is normalized by the cluster virial radius. The numbers in brackets provide numbers of galaxies in each bin for the low-mass hosts first and high-mass hosts second. \emph{Panel (e)}: Distribution of the satellite dwarfs in projected phase-space at $z=0$, where clustercentric distance is normalized by the cluster's virial radius $r_{\rm 200c, host}$ and the line-of-sight velocity $|v_{\text{LOS}}|$ is normalized by a proxy for the cluster's velocity dispersion, the virial velocity $\sqrt{G\MHOST/R_{\rm 200c, host}}$. The phase-space is divided into five regions according to \citet{Rhee2017}, where regions A and E are dominated by first-infallers and ancient-infallers, while the other three regions are composed of varying proportions of first-, recent-, intermediate- and ancient-infallers. \emph{Panel (f)}: Cumulative SFHs for satellite dwarfs in different phase space regions, as defined in panel (e) in two bins of stellar mass (solid and dashed curves). The numbers in brackets provide numbers of galaxies in each bin for the low-mass satellites first and high-mass satellites second.} \label{fig:satelliteEnvironment}
\end{figure*}

\subsubsection{Radial distance from the host centre}

In Fig.~\ref{fig:satelliteEnvironment} (a) \& (b), we show the cumulative SFHs for satellites in a low (top) and high (bottom) bin of stellar mass and host mass (dashed and solid curves), further separated into bins of $z=0$ cluster-centric distance normalized by the cluster's $z=0$ virial radius. The cumulative SFHs of low-mass dwarfs show a noticeable correlation with $z=0$ cluster-centric distance, with dwarfs found closer to their host centre having assembled their stellar mass earlier. This is also true for the more massive dwarfs, but only in more massive hosts. 

Even at distances of $d/r_{\text{200c}}=1-2$, on average, low-mass dwarfs assembled essentially the entirety of their stellar mass over 6~Gyr ago in higher mass clusters ($\MHOST=10^{14-14.3}\MSUN$). This would indicate that most low mass dwarfs, even beyond the virial radii of their host clusters, were quenched several Gyr ago, with those found at $z=0$ closer to their host centre having done so at earlier times. These results are consistent with several previous studies which show that the influence of massive clusters extents to well beyond their virial radii \citep[e.g. see][]{Balogh2000,Hansen2009,vonDerLinden2010,Bahe2013,Wetzel2014,Zinger2018}. The exception to this trend are high mass dwarfs ($\MSTAR=10^{9-10}\MSUN$), which are likely to be more resilient to environmental quenching, or satellites in low-mass hosts ($\MHOST=10^{12-12.5}\MSUN$), where environmental effects are weaker.

\subsubsection{Accretion time} \label{sec:accTime}

Similarly, the time at which the dwarfs were accreted onto their host clusters plays a significant role in determining their cumulative SFH, as seen in Fig.~\ref{fig:satelliteEnvironment} (c) \& (d). We define the accretion time of the galaxy as the last snapshot before the galaxy becomes a part of its $z=0$ host, as defined by \textsc{subfind}, with no restriction placed on cluster-centric distance.\footnote{Note that since the merger trees trace the evolution of the subhaloes, but not the FOF haloes, this procedure assumes that the $z=0$ central has always been the central of its FOF group.}

In Fig.~\ref{fig:satelliteEnvironment} (c), the low mass dwarfs show a dependence on accretion time, regardless of host mass, such that dwarfs accreted earlier assembled their stellar mass earlier. In fact, over half the dwarfs that were accreted more than 4 (8)~Gyr ago onto low mass hosts were quenched more than 4 (7)~Gyr ago. The case is even stronger for dwarfs in high-mass clusters, with the majority of dwarfs accreted over 4~Gyr ago being quenched. These results are consistent with those presented in fig. 4 of \citet[][]{Donnari2020a}, and further extend those results to lower galaxy masses and to TNG50. For the more massive dwarfs (bottom right panel), this trend is also seen to an equal degree in the most massive clusters, but is much weaker in lower mass hosts, although the dwarfs assemble their stellar mass at later times compared to analogous low-mass dwarfs.

\subsubsection{Location in the phase-space plane}
From Fig.~\ref{fig:satHostMass} and Figs.\ref{fig:satelliteEnvironment} (a-d), it is clear that the key driving factor in determining the cumulative SFHs of the satellite dwarfs in our sample is a combination of stellar mass and host mass followed by the radial position of the satellites within their host and the time they have spent as part of the host. In fact, the distance from the host centre and the time since accretion are correlated quantities, as simulations have shown that satellites that fell into their current hosts a longer time ago are typically found closer to their host centres \citep{Rhee2017}. While the projected radial distance of a satellite from its host's centre (and often even the 3D distance with redshift measurements) is readily available in observations, extracting the accretion time is not an easy task. As demonstrated by \citet{Rhee2017} however, the location of satellites on a phase-space diagram can provide a good proxy for their accretion times. Therefore, in Fig.~\ref{fig:satelliteEnvironment} (e), we show the distribution of our satellite dwarfs in projected phase-space, where the line-of-sight velocity $|v_{\text{LOS}}|$ is measured along a random orientation (i.e. along the \textit{z}-axis), and we use it to separate our entire sample of satellite dwarfs into five regions in projected phase-space following \citet{Rhee2017}. The radial distance is normalized by the host's virial radius $r_{\rm 200c, host}$ and to normalize $|v_{\text{LOS}}|$, we use a proxy for the cluster's velocity dispersion, the virial velocity $\sqrt{G\MHOST/R_{\rm 200c, host}}$. We have confirmed that ancient-infallers ($t_{\text{acc,lookback}}>8$ Gyr ago) are concentrated in regions E and D, and to some extent, C, while the intermediate-infallers ($t_{\text{acc,lookback}}=4-8$ Gyr ago) have a similar but broader distribution, extending into regions B and A as well. Recent-infallers ($t_{\text{acc,lookback}}<4$ Gyr ago) on the other hand are nearly uniformly distributed in the regions D and C and the adjoining parts of region A. The corresponding cumulative SFHs for satellites in the different phase-space regions are shown in Fig.~\ref{fig:satelliteEnvironment} (f), for the same two bins of stellar mass. 

At both low and high stellar masses, satellites in region E, i.e. mostly ancient-infallers, show the earliest SFHs, which those in region A, i.e. mostly first-infallers, show the latest SFHs. Note that unlike \citet{Rhee2017}, region A here does not contain any interlopers, since all satellites are associated with their FOF host haloes, although $\sim0.8\%$ can be found a distances $>3R_{\rm 200c, host}$. Dwarfs in regions B, C and D have intermediate SFHs, consistent with those subsamples having varying proportions of ancient-, intermediate-, recent- and first-infallers. In fact, these results may also be interpreted in the context of the variation of tidal forces with both radial position and velocity, defining the orbit of the galaxies within the host halo \citep[e.g. see][]{Gnedin1999}. These tidal forces are likely to indirectly result in the quenching of the galaxies, giving rise to the dependence on phase space we have seen in Fig. \ref{fig:satelliteEnvironment} (f). We have confirmed that in narrower bins of host mass, these trends hold qualitatively, although the shapes of the cumulative SFHs may differ. In fact, it is always the case that dwarfs in region E have earlier SFHs compared to those in region A; the SFHs in regions B, C and D do not always show the same monotonic trends, but any deviations are largely due to low-number statistics.

\subsection{Impact of local environment} \label{sec:localEnv}

\begin{figure*}
 \includegraphics[width=\linewidth]{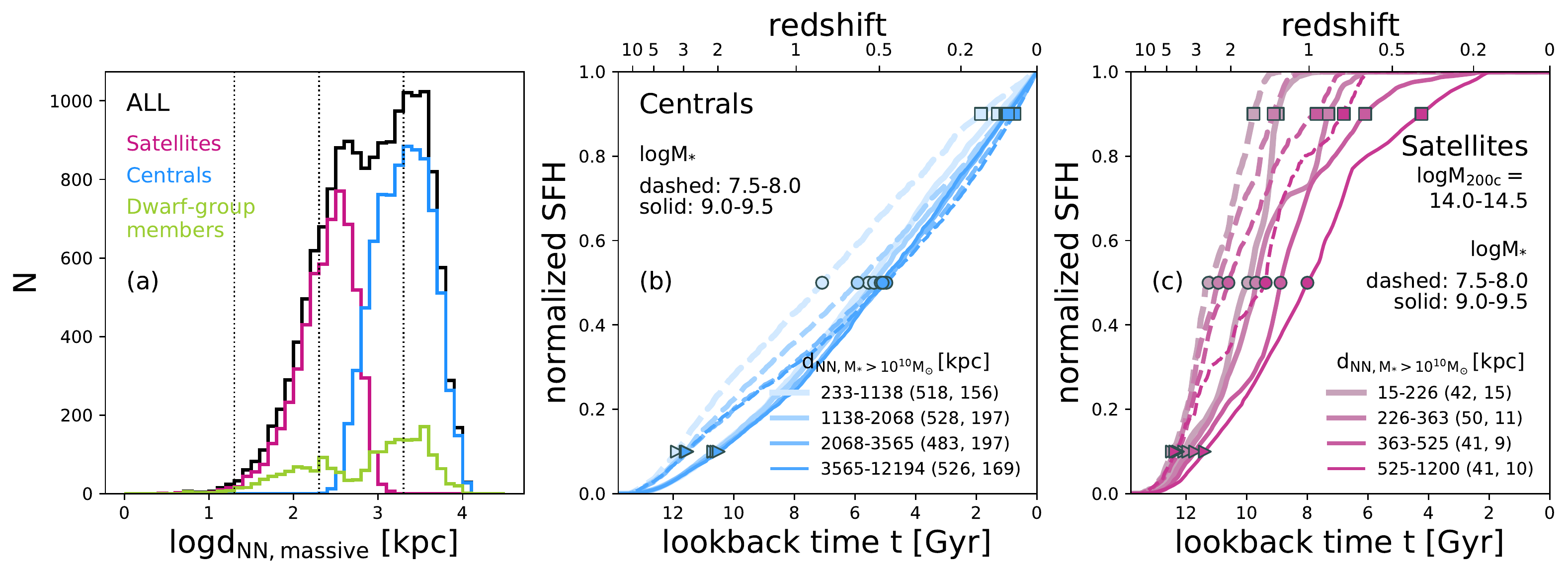}
 \caption{\textbf{Dependence on local environment of TNG50 dwarfs.} \emph{(a)}: Distribution of distance to nearest massive neighbour i.e. with $\MSTAR>10^{10}\MSUN$. Vertical dotted lines mark 20, 200 and 2000 kpc for reference. \emph{(b) \& (c)}: Cumulative SFHs for centrals (panel (b)) and satellites (panel (c)) in bins of distance to nearest massive neighbour, restricted to a low (dashed curves) and high (solid curves) stellar mass bin. The satellites are also further restricted to the most massive host mass bin of $M_{\text{200c}}=10^{14-14.3}\MSUN$. The bins are the quartiles for the full sample of central and satellite dwarfs respectively. The numbers in brackets provide numbers of galaxies in each bin for the low-mass hosts first and high-mass hosts second.} \label{fig:nearestNeighbour}
\end{figure*}

While the mass of the host group/cluster is an important indicator of environment, dwarfs in these host clusters can still be found in a large range of densities, especially in the case of the larger clusters. Furthermore, central dwarfs may also be affected by their local environment, even though they may not be orbiting within massive host haloes. We hence here explore another measure of environment, $d_{\text{NN,massive}}$ i.e. the distance to the nearest massive neighbour i.e with $\MSTAR>10^{10}\MSUN$. 
In Fig.~\ref{fig:nearestNeighbour} (a), we show the distribution of $d_{\text{NN,massive}}$, for the full sample of dwarfs as well as for the satellite, central and dwarf group subsamples, for context. Satellites, centrals and dwarf group members occupy different ranges of distance to their nearest massive neighbour: $d_{\text{NN,massive}}\simeq3-1500$, $230-12200$, and $5-12200$ kpc, respectively. In the case of the satellites, the nearest massive neighbour is likely to be within the same FOF group as the satellites themselves. For the centrals, which are usually (but not necessarily) the most massive galaxy in their halo, the nearest massive neighbour is likely to be outside the FOF group. In the case of the dwarf group galaxies, we see a double-peaked distribution, where the peak at smaller $d_{\text{NN,massive}}$ probably represents systems containing an $\sim L_{*}$ central galaxy with a system of dwarfs, whereas the peak at larger distances is likely to be groups of dwarfs, where the nearest massive neighbour is outside the FOF group they belong to.

In Fig.~\ref{fig:nearestNeighbour} (b) \& (c), we show the impact of local environment separately for the centrals and satellites, further controlling for the host mass in the case of the satellites, and to understand whether the local environment actually has an effect on all subsamples of the dwarf population. We again focus on two galaxy stellar mass ranges, $\MSTAR=10^{7.5-8}\MSUN$ and $\MSTAR=10^{9-9.5}\MSUN$. The satellites are further restricted to a narrow range in host mass, $\MHOST=10^{14-14.3}\MSUN$ in addition to the stellar mass bins. There is little dependence on local environment for the centrals (except for the low mass centrals in the two densest quartiles of environments) as well as the low mass satellites. Only the most massive satellites show significant residual dependence on local environment, but note that this is not the case in less massive hosts (not shown). Additionally, since we do not exclude the central galaxy of the host cluster from being considered for the nearest most massive neighbour for the satellites, this dependence is likely a reflection of the dependence on cluster-centric radial position rather than the presence of other nearby massive galaxies.

From these results, it is clear that the local environment, as quantified by the distance to the nearest massive neighbour, is at best a secondary factor in shaping the cumulative SFHs of the dwarfs. The dominant factors are the satellite/central status of the dwarfs, their stellar mass, and their host mass for satellites.

%-------------------------------------------------

\section{Discussion} \label{sec:discussion}

\subsection{The connection between quenching, SFHs and $\tau_{90}$} \label{sec:discQuenching}

\begin{figure*}
    \includegraphics[width=\linewidth]{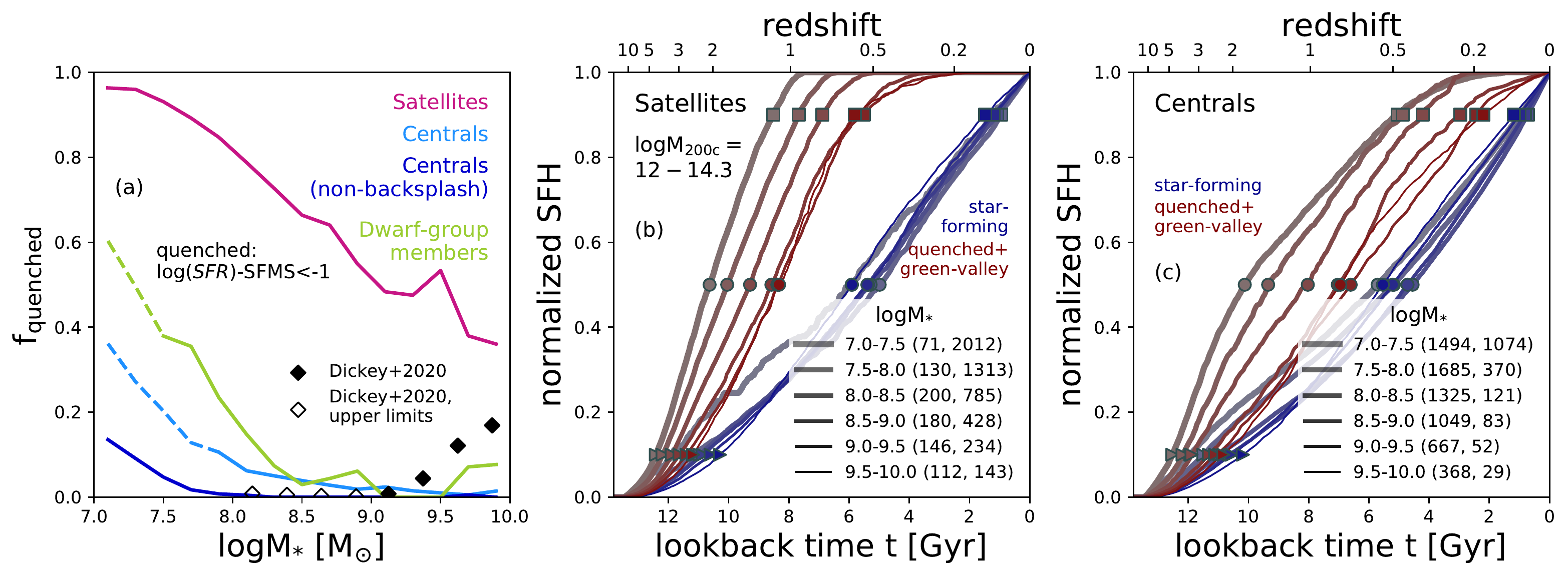}
    \includegraphics[width=\linewidth]{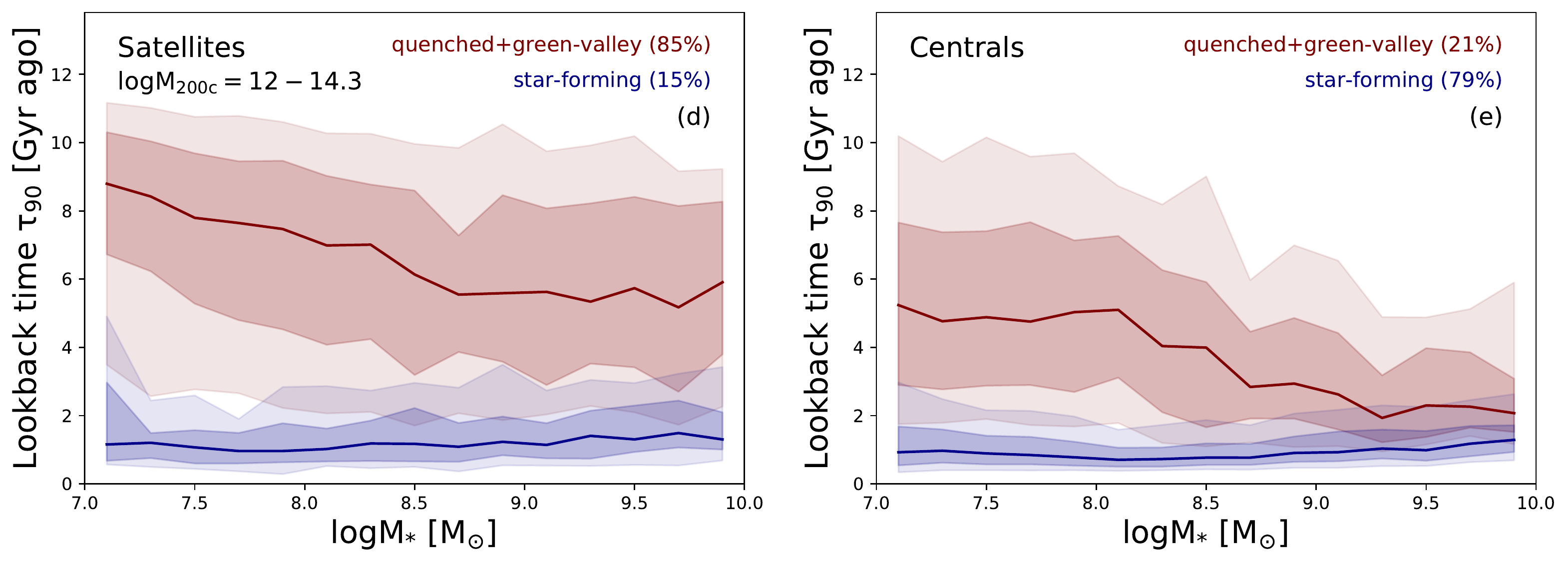}
    \caption{\textbf{Connection between shapes of SFHs and quenching status at $z=0$} \emph{(a)}: Fractions of satellites, centrals and dwarf group members that are quenched at $z=0$, as a function of stellar mass. We do not include galaxies that are in the green valley i.e. with $-1<\log{({\rm SFR})}-{\rm SFMS}<-0.5$, but note that these represent 4-7.5\% of each of the three subsamples. Dashed curves indicate mass regimes where the quenched fractions are known to be overestimated due to resolution effects. See text for details. Additionally, with the dark blue curve, we show the quenched fractions for centrals that are non-backsplash galaxies (here, chosen as those centrals that were not satellites at any time since $z=2$.) For comparison, we show the quiescent fractions for isolated SDSS galaxies by \citet{Geha2012}, as reported in \citet{Dickey2020}, as black points; open points indicate upper limits where the number of isolated quiescent galaxies is zero in their sample. \emph{(b) \& (c)}: Cumulative SFHs for satellites and centrals, separately for quenched (maroon) and star-forming (blue) galaxies, in bins of stellar mass. The first and second number in each bracket indicate the number of galaxies in the given mass bin that are star-forming and quenched respectively. \emph{(e) \& (f)}: Distribution of critical assembly times, $\tau_{90}$, as a function of stellar mass, for satellites and centrals, separately for star-forming (blue) and quenched (quenched) galaxies. The curves indicate the median $\tau_{90}$ values in 0.25 dex bins of stellar mass, while the darker shaded areas indicate the $20^{\rm th}-80^{\rm th}$ percentile scatter and the lighter areas, the $5^{\rm th}-95^{\rm th}$ percentile scatter.} \label{fig:sfVsQuenched}
 \end{figure*}

\begin{figure*}
    \includegraphics[width=\linewidth]{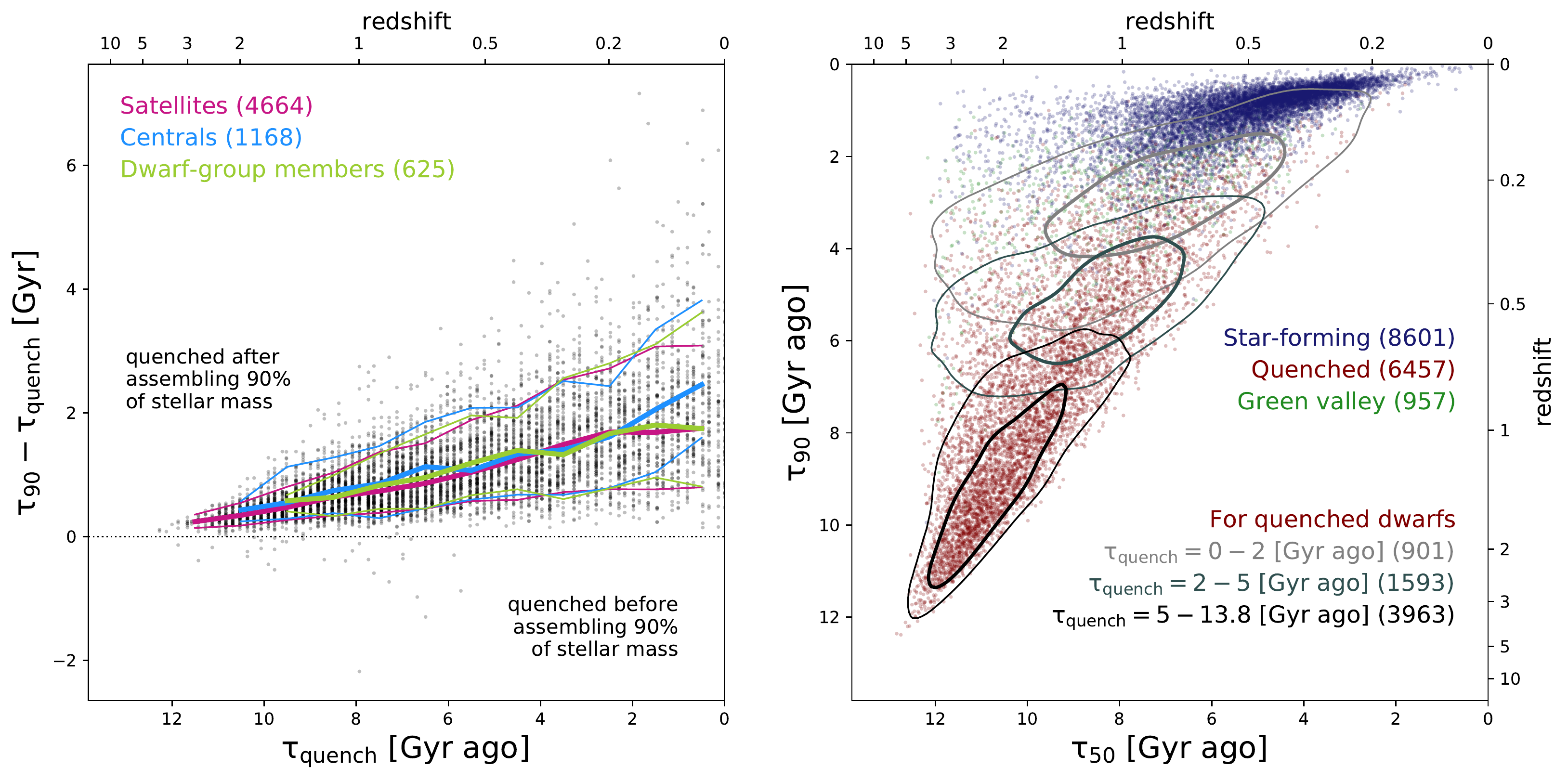}
    \caption{\textbf{Connection between $\tau_{50}$, $\tau_{90}$ and $\tau_{\mathrm{quench}}$.} \emph{Left}: Difference between $\tau_{90}$ and $\tau_{\mathrm{quench}}$ (both in the form of lookback times) as a function of $\tau_{\mathrm{quench}}$ for all quenched dwarfs. Positive (negative) values of $\tau_{90}-\tau_{\mathrm{quench}}$ indicate that quenching occurred after (before) the galaxy has assembled 90\% of its stellar mass. Thick coloured curves show the median values of $\tau_{90}-\tau_{\mathrm{quench}}$ in 1~Gyr bins of $\tau_{\mathrm{quench}}$, for the satellite, central and group dwarf subsamples; thin curves indicate the corresponding $10^{\rm th}-90^{\rm th}$ percentile values. \emph{Right}: Loci of star-forming, quenched and green-valley dwarfs in the $\tau_{90}-\tau_{90}$ plane. For the $z=0$ quenched dwarfs only (maroon points), contours show the loci for dwarfs quenched within a given epoch (thick and thin contours indicate 50\% and 90\% of the subsample). The number in each bracket indicates the size of the subsample.}
    \label{fig:t90VstQuench}
\end{figure*}

In Section~\ref{sec:SFHs}, we have shown that most $z=0$ central dwarfs reach $\tau_{90}$ within the last $\sim0.5-2$~Gyr throughout the $10^{7-10}\MSUN$ stellar mass range, with little scatter. On the other hand, $z=0$ satellite dwarfs in hosts with masses of $\MHOST\gtrsim 10^{12}\MSUN$ have assembled most of their current mass by at least $z\sim0.5$, i.e. at least 6 billion years ago, but with large galaxy-to-galaxy variations depending on galaxy stellar mass, current host mass, location, and accretion time (see Section~\ref{sec:environmentalFactors}). In Section~\ref{sec:TNG50dwarfs}, we have also shown that the majority of TNG50 satellites in massive hosts are red galaxies. In fact, both in the low-redshift Universe \citep{Geha2012} and in the TNG simulations \citep[e.g.][]{Donnari2020a}, low-mass galaxies are rarely quenched (with quenched fraction $\lesssim 10$ per cent) \emph{unless} they are satellites of more massive galaxies, i.e. unless they are not isolated. For individual dwarfs, the cumulative SFHs have markedly different shapes depending on whether they are quenched, and the time at which they quenched, or if they are still star-forming. The stacked cumulative SFHs are similarly shaped by the proportion of star-forming to quenched galaxies in the subsamples. We delve more explicitly into this connection in what follows.

\subsubsection{Cumulative SFHs and assembly times for quenched and star-forming galaxies}

We use the star-formation activity flags of \citet{TNG50Pillepich2019} to determine the star-forming status of TNG50 galaxies, so that galaxies whose specific star formation rates (sSFRs) are 1 dex below the corresponding star-forming main sequence (SFMS) are deemed to be `quenched', those with $\log{\rm sSFR}-$SFMS$ >-0.5$ dex are `star-forming' and those with intermediate values are considered to be in the `green-valley'. Here, the SFMS is determined recursively (see \citealt{TNG50Pillepich2019} or \citealt{Donnari2020b} for details). 

Fig.~\ref{fig:sfVsQuenched} (a) shows the fractions of TNG50 dwarfs that are quenched at $z=0$ as a function of galaxy stellar mass, using this method. The results from TNG50 are shown as dashed curves in those regimes where we know that the finite mass resolution of TNG50 is responsible for a systematic overestimate of the quenched fractions of more than 10 percentage points -- see Appendix~\ref{sec:appResEffects}: this occurs at $\MSTAR\lesssim10^{7.5-8}\MSUN$ at TNG50 resolution for dwarf-group members and central galaxies, while for satellites our resolution convergence tests do not raise any quantitative concern. Satellite dwarfs of massive hosts are more frequently quenched than dwarf-group members, and much more than dwarf centrals. We show results from \citet{Dickey2020} with black data points in Fig.~\ref{fig:sfVsQuenched} (a) for comparison, although note that this is done at face-value without any mocking of the simulation data. The TNG50 results appear to underestimate the quenched fractions at masses $\MSTAR=10^{9-10}\MSUN$ by up to 5-8 percentage points and to overestimate them by up to a few percentage points at $\MSTAR=10^{8-9}\MSUN$. However, the interested reader should refer to \citet{Dickey2020} and \citet{Donnari2020b} for more rigorous comparisons and Appendix~\ref{sec:appResEffects} for additional comments on the effects of numerical resolution on the quenched fractions.

According to TNG50, the quenched fraction is less than 5 per cent and approximately constant for centrals of mass $\MSTAR=10^{8.5-10}\MSUN$. However, we have verified that $\sim15-35$ per cent of the centrals, depending on their stellar mass, are in fact backsplash galaxies, defined here to mean centrals that were satellites for some time between present day and $z=2$, based on the classifications of \citet{Zinger2020}. These account for a significant portion of the quenched centrals (see also \citealt{Zinger2020} for a similar discussion with the TNG100 and TNG300 simulations). Excluding such galaxies, the quenched fractions for the \emph{non-backsplash} centrals are also shown in Fig.~\ref{fig:sfVsQuenched} (a) with the dark blue curve. The plot shows that when we exclude backsplash galaxies, the quenched fraction is practically zero for stellar masses of $\MSTAR=10^{8-10}\MSUN$ and at most 14 per cent at lower masses in our sample. The effects of resolution mentioned above and quantified in Appendix~\ref{sec:appResEffects} are likely also partially due to backsplash galaxies. Additionally, it should be noted that the declining quenched fraction of \emph{satellites} with stellar mass is largely dominated by low-mass hosts, ($f_{\text{quenched}}$ decreases from 91 to 2 per cent in $\MHOST=10^{12-12.5}\MSUN$ hosts); in the most massive hosts ($\MHOST=10^{14-14.3}\MSUN$ hosts), on the other hand, the quenched fractions have a milder trend with stellar mass, in agreement with previous findings based on both the TNG and other numerical simulations (see e.g. fig.10 upper right panel of \citealt{Donnari2020b}).

The cumulative SFHs for the satellites and centrals -- as in the lower panels of Fig.~\ref{fig:cumulSFHsAndMstarBins} -- are shown in Figs.~\ref{fig:sfVsQuenched} (b) \& (c), now further separated by whether they are quenched or star-forming at $z=0$. It is clear that the star-forming dwarfs, whether they are satellites or centrals, have similarly shaped SFHs, which also have little dependence on stellar mass. This is not the case, however, for quenched (or green-valley) galaxies. In the case of the satellites, the cumulative SFHs suggest a near constant average SFR until the galaxies are quenched, with median quenching redshifts ranging from $\sim0.2-1$. There is a mild dependence on stellar mass, with the values of $\tau_{90}$ differing by $\sim1$~Gyr for every 0.5 dex in stellar mass, up to $\MSTAR=10^{9}\MSUN$; at higher masses, the SFHs are nearly identical to the satellites with $\MSTAR=10^{8.5-9}\MSUN$. For the centrals at the low mass end ($\MSTAR=10^{7-8}\MSUN$, the SFHs suggest higher average SFRs at $z\gtrsim2$, and then reduced SFRs at later times, ultimately quenching at median redshifts of $\sim0.2$. At higher masses, the SFHs for quenched centrals are similar, and extend nearly to present day.

Figs.~\ref{fig:sfVsQuenched} (d) \& (e) similarly show the dependence of $\tau_{90}$ on stellar mass for quenched and star-forming galaxies separately and prompt similar conclusions: the $\tau_{90}$ values are similar for star-forming satellites and star-forming centrals at all masses, with an approximate value of 1~Gyr ago. The quenched/green-valley galaxies, on the other hand, show a mildly decreasing trend of $\tau_{90}$ within the mass range of $\MSTAR\sim10^{7-8.5}\MSUN$, from $\sim9$~Gyr ago to $\sim6$~Gyr ago for satellites and from $\sim6$~Gyr ago to $\sim2$~Gyr ago for centrals. At higher masses, the trend is approximately constant. Note that if we consider the green-valley galaxies separately, they also display approximately constant values of $\tau_{90}\sim2-3$~Gyr for satellites and centrals of all masses.

Hence, the trends seen in Fig.~\ref{fig:t50t90AssemblyTimes} are a convolution of the individual trends for quenched and star-forming galaxies and the dependence of quenched fractions on stellar mass. These results also show that even those centrals that are quenched at $z=0$, were quenched more recently, compared to the satellites of the same mass.

\subsubsection{Mapping between time since quenching and $\tau_{90}$}

While the values of $\tau_{90}$ have been broadly used throughout the community as proxies for the quenching times of galaxies \citep[e.g.][]{Weisz2019}, such a correspondence only applies for galaxies that are actually quenched at $z=0$. And even in those cases, the time when 90 per cent of the final $z=0$ stellar mass is assembled does not necessarily reflect the time when a galaxy actually stopped forming stars. This is also the case for dwarf galaxies, where the contribution to the stellar mass via accretion (ex-situ stars) is negligible. Hence, here we quantify the mapping between the stellar assembly time $\tau_{90}$ and the time since quenching, by directly measuring the latter in the simulation.

We define $\tau_{\mathrm{quench}}$ as the time at which a galaxy last transitioned from star-forming/green-valley to quenched (i.e. had $\log{\rm sSFR} <$ SFMS$-1.0$ dex) and remained quenched through $z=0$. The left panel of Fig.~\ref{fig:t90VstQuench} shows the difference between $\tau_{90}$ and $\tau_{\mathrm{quench}}$ as a function of $\tau_{\mathrm{quench}}$ itself, where both quantities are in the form of lookback times. Note that for 7 of the 6500 quenched dwarfs in TNG50, we were unable to measure a quenching time, so these dwarfs are excluded from this analysis. The data points in Fig.~\ref{fig:t90VstQuench} represent 4664, 1168, and 625 \emph{quenched} satellites, centrals, and dwarf-group members respectively.

A few key features are evident in Fig. \ref{fig:t90VstQuench}: as expected, nearly all quenched dwarfs quench after assembling 90 per cent of their stellar mass. For galaxies that quenched at very early times, i.e. $\tau_{\mathrm{quench}}>11$~Gyr ago, by design there is little difference between the two quantities. For later quenching times however, there is a consistent trend of increasing scatter in $\tau_{90}-\tau_{\mathrm{quench}}$ with decreasing $\tau_{\mathrm{quench}}$; the median $\tau_{90}-\tau_{\mathrm{quench}}$ rises from $\sim0.35$~Gyr at $\tau_{\mathrm{quench}}=10$ to $\sim1.9$~Gyr at $z=0$. There are no significant differences between satellites, centrals or group dwarfs, nor between dwarfs of different masses. The $1\sigma$ scatter in the values of $\tau_{90}-\tau_{\mathrm{quench}}$ also rises with decreasing $\tau_{\mathrm{quench}}$ from $\sim0.36$~Gyr at $\tau_{\mathrm{quench}}=10$ to $\sim2$~Gyr at $z=0$. These results imply that, on average, for galaxies that quenched around $z\sim1-2$, quenching can occur up to $0.2-1$~Gyr after 90 per cent of their stellar mass has been assembled; on the other hand, for galaxies that quenched more recently, at $z\lesssim1$, actual quenching can occur on average (for the 90th percentiles) after about 2 (3) billion years after the assembly of 90 per cent of the final stellar mass. These numbers can be used as guidelines for the systematic errors that can be incurred when using $\tau_{90}$ as a proxy for quenching time\footnote{Note that we have examined the 21 satellites and 8 centrals which quench before assembling 90\% of their stellar mass and find no unusual or biased SFHs for these dwarfs, with the exception that all of the centrals have masses $\MSTAR=10^{7-8}\MSUN$.}. 

Finally, the right panel of Fig.~\ref{fig:t90VstQuench} shows the loci occupied by TNG50 dwarfs that at $z=0$ are star-forming, green-valley and quenched galaxies in the $\tau_{90}$ vs. $\tau_{50}$ plane. The location of a galaxy on this plane gives some indication of the shape of its SFH i.e. how relatively rapidly did it assemble the first half of its stellar mass compared to the second half. Furthermore, since the stellar assembly times $\tau_{90}$ vs. $\tau_{50}$ are more readily available, e.g. from the SFHs of the galaxies, compared to the accretion times, this plot allows us to approximately map the latter to the former. By construction, all galaxies are above the 1:1 line. Star-forming galaxies can only be found in the upper/upper-right portion of the plot, with very recent assembly times for both the 50 and, even more so, the 90 per cent of the final stellar mass. On the other hand, quenched galaxies occupy a much wider region of the $\tau_{90}$ vs. $\tau_{50}$ plane, depending on the time since quenching: see gray contours. Following \citet{Weisz2019}, in an upcoming paper we will use this diagnostic plot to contrast the satellite populations of MW-like and Andromeda-like satellites emerging from TNG50.

\subsection{Enhanced star-formation of satellites at early epochs? A subtle environmental effect prior to accretion}

\begin{figure*}
    \includegraphics[width=\linewidth]{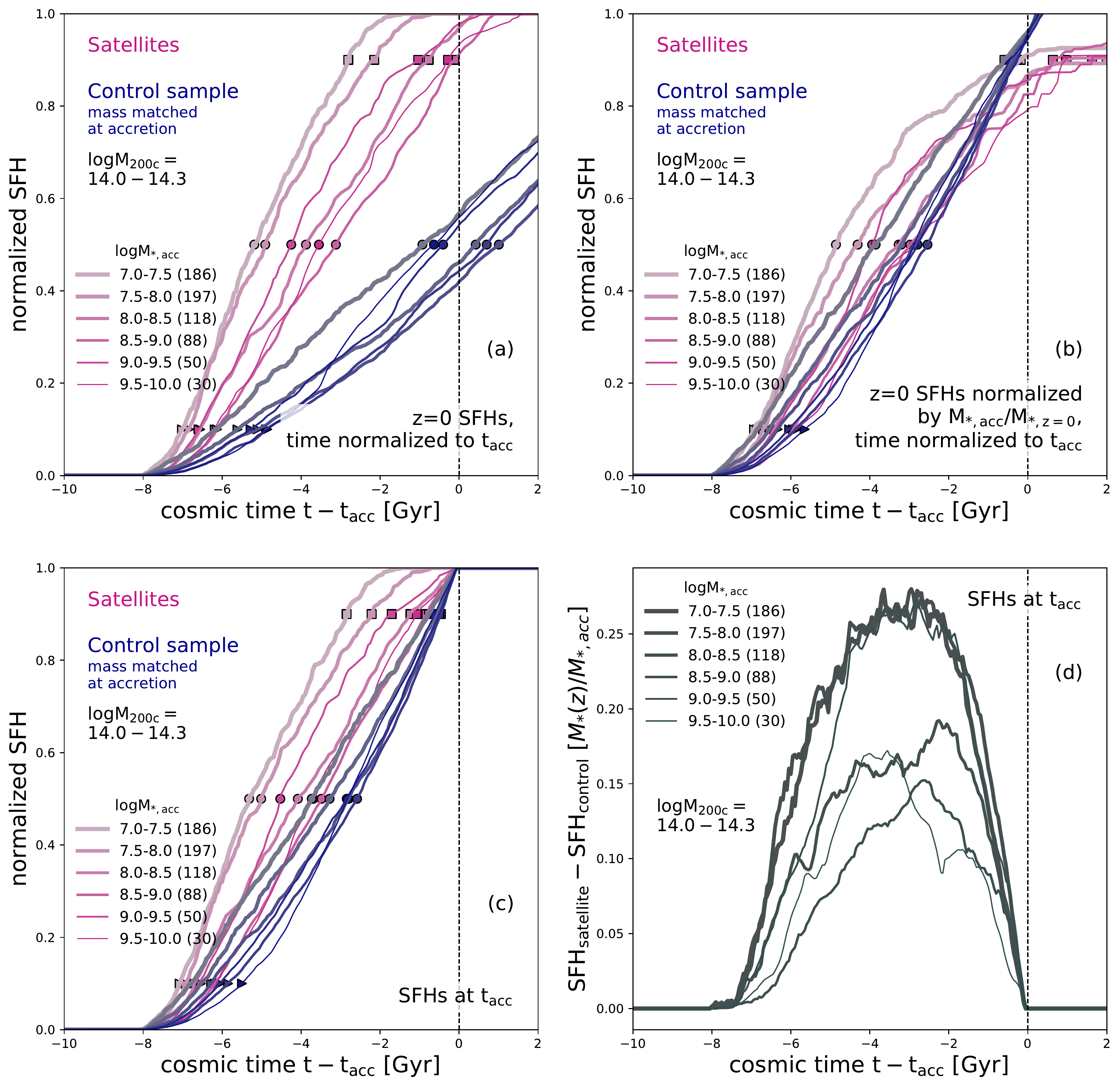}
    \caption{\textbf{Comparison between TNG50 satellite and control SFHs to quantify whether satellites exhibit more efficient SFHs at early times.} \emph{Panels (a), (b) and (c)}: Cumulative SFHs of satellites (pink) restricted to hosts of mass $\MHOST=10^{14-14.3}\MSUN$ compared to a control sample of galaxies mass-matched at accretion, in bins of stellar mass \emph{at accretion}, with time relative to accretion time $t_{\mathrm{acc}}$. Note that we exclude 4 dwarfs with $M_{\rm *,acc}<10^{7}\MSUN$ and 2 with $M_{\rm *,acc}>10^{10}\MSUN$ for clarity. In panel (a), the SFHs are simply the medians of the $z=0$ SFHs; in panel (b), individual SFHs are first normalized by $M_{\mathrm{*,acc}}/M_{\mathrm{*,z=0}}$ before calculating the medians; in panel (c), the SFHs were calculated using stellar particles belonging to the subhalo at $t_{\mathrm{acc}}$. \emph{Panel (d)}: Difference between the median cumulative SFHs of satellites and the corresponding control sample from panel (c). The number in each bracket indicates the size of the subsample.} \label{fig:earlySFSatVsControl}
\end{figure*}

Throughout the preceding analysis, the satellite SFHs have been shown to systematically tend towards earlier stellar mass assembly when compared to analogous centrals, with somewhat \emph{truncated} cumulative SFHs due to quenching. Given that we always compare satellites and centrals that have similar stellar masses at $z=0$, this also implies that satellites may be more efficient at making stars (i.e. have higher SFRs) at early epochs, while centrals are more efficient at later times, when the satellites have been accreted by their host haloes and are subject to various environmental effects. A natural question to ask is whether these differences only stem from the changes to the satellites' behaviour after accretion, or whether they also have enhanced SFRs at very early epochs, well \emph{before} accretion onto the hosts where they are found at $z=0$.

In order to directly compare the satellites' SFHs before accretion to those of analogous centrals, with TNG50 we can select a control sample of central galaxies that are mass-matched to the satellites at their time of accretion, which remain centrals until $z=0$. As mentioned in Section \ref{sec:accTime}, we define the accretion time of the galaxy as the last snapshot before the galaxy becomes a part of its $z=0$ host (and not when it crossed $R_{\text{vir}}$). For this section, the snapshot at which the control galaxy was matched to a satellite is referred to as its `accretion' snapshot for simplicity. We first directly compare the SFHs of the satellites and control samples, with time relative to this accretion time in Fig.~\ref{fig:earlySFSatVsControl} (a). Only results for satellites in the most massive hosts ($\MHOST=10^{14-14.3}\MSUN$) are shown in the figure since any trends are strongest in this host mass regime. The galaxies are binned by their stellar mass \emph{at accretion} (note that we only select the satellites by their properties and simply consider their control counterparts). It is clear that well before the time of accretion, the satellites have assembled significantly more mass than the control galaxies at all masses. However, this is to be expected, since the satellites show little growth in stellar mass after accretion, while the control galaxies continue to grow in stellar mass to $z=0$; therefore, although the satellites and control galaxies have similar masses at accretion, these masses represent significantly different proportions of their final stellar masses and hence their cumulative SFHs. 

In order to determine if the satellites are truly more efficient at early times, we therefore renormalize the cumulative SFHs of the individual satellites and control dwarfs by $M_{\rm *,acc}/M_{\rm *,z=0}$ and show the median results in panel (b) of Fig.~\ref{fig:earlySFSatVsControl}. The figure now shows that the SFHs of the satellites do indicate earlier assembly of the mass at accretion. However, in principle, the renormalized cumulative SFHs of all galaxies should equal 1 at $t=t_{\mathrm{acc}}$. This is not the case due to the fact that the galaxies may lose stellar mass through processes such as tidal stripping, as well as gain particles through mergers or accretion, which results in $M_{\rm *,acc}/M_{\rm *,z=0}$ not being equal to the cumulative SFH at $t=t_{\mathrm{acc}}$. As these processes occur after $t=t_{\mathrm{acc}}$, simply renormalizing the $z=0$ SFHs is not sufficient to truly compare the SFHs of the satellites and control dwarfs before accretion. 

We have therefore measured new SFHs for both the satellites and control galaxies using the stellar particles they contain at the accretion snapshot and compare the results in panel (c) of Fig~\ref{fig:earlySFSatVsControl}. It is clear from this figure that the satellites do indeed assemble their stellar mass at accretion earlier than the control galaxies at all masses. These differences are shown explicitly in panel (d) of Fig.~\ref{fig:earlySFSatVsControl}, which shows that the peak difference between the median cumulative SFHs is at least $\sim0.15$ and can be as high as $\sim0.3$, with the peak difference occurring $\sim2-4$~Gyr before the time of accretion, $t_{\mathrm{acc}}$. In other words, as of $\sim2-4$ billion years before accretion, $t_{\mathrm{acc}}$, the satellites-to-be of hosts with $\MHOST=10^{14-14.3}\MSUN$ at $z=0$ have assembled $\sim15-30$ per cent more of their stellar mass at accretion compared to the control sample. We find no monotonic trend with stellar mass for this enhanced SF activity before accretion. Moreover, these trends are strongest for satellites of the most massive hosts and become weaker at lower host mass regimes, negligible at $\MHOST=10^{12-13}\MSUN$ (not shown).

These results should be considered with caution, however, since they are likely to be affected by preprocessing of galaxies in lower mass hosts before first being accreted onto their $z=0$ hosts \citep[see e.g.][]{Donnari2020a}; this is supported by the fact that $\sim35-40$ per cent ($\sim60-65$ per cent) of these $\MHOST=10^{14-14.3}\MSUN$ satellites with mass $M_{\mathrm{*,acc}}=10^{8-10}\MSUN$ ($M_{\mathrm{*,acc}}=10^{7-8}\MSUN$) are already quenched at accretion. It is likely that a large portion of those satellites that have been preprocessed were quenched even before they were accreted onto the final host. Therefore, comparing such galaxies to the control sample is analogous to comparing quenched satellites at $z=0$ to the centrals at $z=0$, the majority of which are star-forming: the preprocessed satellites would appear to have higher SFRs at earlier times as, by design they will have assembled their $t_{\text{acc}}$ stellar mass before quenching (and well before $t_{\text{acc}}$), while the control galaxies do so more gradually up to $t_{\text{acc}}$. If we consider only those satellites that are star-forming at accretion and their control counterparts for the results of Fig.~\ref{fig:earlySFSatVsControl}, we find qualitatively but not quantitatively consistent trends. Namely, the SFHs of satellites-to-be that are star forming at accretion imply earlier stellar mass assembly compared to the star-forming control galaxies only in certain stellar mass bins, with no clear trends. 

Within the aforementioned caveats, the results of Figure~\ref{fig:earlySFSatVsControl} do suggest that a significant portion of the $z=0$ satellites do in fact exhibit enhanced SFRs at early epochs, i.e. well before accretion, in comparison to analogous dwarfs that will not become satellites of massive hosts at future times. The presence and degree of such enhancement depends on the final host mass and whether or not satellites-to-be are quenched or star-forming at accretion (likely due to preprocessing). These results indicate that the large-scale environment of the satellites at very early epochs, and before any host-satellite physical processes start to affect the SFHs, may also be an important factor. This represents a subtle environmental effect, possibly due to the availability of gas in those regions of the Universe that eventually collapse in very massive galaxy clusters at lower redshift. Such an effect may also be understood in the context of assembly bias, whereby galaxy properties are correlated with the properties of their host haloes. However, further, more careful analysis would be required to separate the effects of assembly bias and preprocessing. Within the TNG model, we have also seen indications of a similar early, pre-accretion environmental effect in another context: in relation to the enhanced gas-phase metallicity of satellites of $z=0$ massive hosts, whereby according to the TNG simulations such enhanced enrichment of satellite vs. field galaxies at fixed stellar mass is established in satellites prior to infall into their final cluster potential \citep{Gupta2018}.

\subsection{Modelling the cumulative SFHs} \label{sec:discModel}
In this section, we provide a simple parametric model for the cumulative SFHs predicted by TNG50: this will not only aid in comparing the SFHs of the different subsamples, but also in comparing our simulation results with future results. 

Galaxy SFHs are often described by several well-established functions such as the tau or delayed-tau models, a lognormal or a double-power law models \citep[e.g. see][and references therein]{Carnall2019}. We have attempted to fit these parametric functional forms to the cumulative SFHs of our samples and find that, at least in the case of quenched galaxies, none of them are able to capture the shape of the cumulative SFHs from TNG50 at early and late times (although all except the delayed-tau model do provide reasonable fits at intermediate times i.e. $t\sim2-8$ Gyr). These models predict much more rapid increases in stellar mass at very early times and too shallow a profile at late times compared to our TNG50 results. In the case of the star-forming galaxies, all of the models except the delayed-tau model do in fact produce reasonable fits; however, the tau- and lognormal formulas both require likely unphysical parameters to do so. Therefore, for this analysis, we choose instead to employ a piecewise-defined fitting formula comprised of different, simply-modelled phases. We assume two phases for the star-forming dwarfs and three for the quenched ones:
\begin{enumerate}
    \item A initial phase with a constant ${\rm sSFR}=\alpha$, which accounts for the rapid, nearly exponential growth of $m(t) = M(t)/M_{*,z=0}$ at early times.
    \item A second phase with a constant ${\rm SFR}=M_{*,z=0} \times \beta$, which results in linear growth of $m(t)$ with cosmic time.
    \item A final phase with an exponentially declining ${\rm SFR}=M_{*,z=0} \times\nolinebreak \beta e^{-(t-t_{2})/\tau}$.
\end{enumerate}

The first two phases are common to both star-forming and quenched galaxies, while the third is only applicable to quenched galaxies. By integrating the piecewise-defined SFRs and ensuring continuity at the transition times, we obtain the following functional form for star-forming galaxies:
\begin{equation} \label{eq:modelSF}
m(t) = 
\begin{cases} 
    m_{0}e^{\alpha t} &; \quad t<t_{1} \\
    m_{0}e^{\alpha t_{1}}\left[\alpha (t-t_{1}) + 1\right] &; \quad t_{1}<t<t_{z=0}
\end{cases}
\end{equation}
with 
\begin{equation} \label{eq:m0SF}
    m_{0} = \frac{1}{e^{\alpha t_{1}}\left[\alpha(t_{z=0}-t_{1}) + 1\right]}
\end{equation}
and the following functional form for quenched galaxies:
\begin{equation}
m(t) = 
\begin{cases}  \label{eq:modelQuenched}
    m_{0}e^{\alpha t} &; \quad t<t_{1} \\
    m_{0}e^{\alpha t_{1}}\left[\alpha (t-t_{1}) + 1\right] &; \quad t_{1}<t<t_{2}\\
    \begin{split}
    & m_{0}e^{\alpha t_{1}}\times \\ & \quad\Bigl[\alpha\tau\left(1-e^{-(t-t_{2})/\tau}\right) \\ & \quad+ \alpha (t_{2}-t_{1}) + 1\Bigr]
    \end{split} &; \quad t_{2}<t<t_{z=0}\\
\end{cases}
\end{equation}
with 
\begin{equation} \label{eq:m0Quenched}
    m_{0} = \frac{1}{e^{\alpha t_{1}}\left[\alpha\tau\left(1-e^{-(t_{z=0}-t_{2})/\tau}\right)+ \alpha(t_{z=0}-t_{1}) + 1\right]}
\end{equation}
where all times here are cosmic times and the value of $m_{0}$ is determined by requiring that $m(t_{z=0})=1$. Additionally, for both star-forming and quenched galaxies, the value of $\beta$ is set by the continuity requirement at $t=t_{1}$, such that $\beta = \alpha m_{0}e^{\alpha t_{1}}$. 

Thus the best-fitting model for star-forming galaxies has two free parameters ($\alpha$ and $t_{1}$) and for quenched galaxies has four (  $\alpha$, $t_{1}$, $t_{2}$ and $\tau$). Note that the model for the star-forming galaxies, specifically for low- and intermediate-mass ($\MSTAR=10^{7-9}\MSUN$) TNG50 centrals, overestimates the cumulative SFHs at early times i.e. $z\gtrsim2$. This indicates that perhaps star-forming galaxies require more complexity for the accurate description of their SFHs. We will explore this in more detail in an upcoming companion paper. In Appendix~\ref{sec:appModelFits}, we provide the best-fit values for these parameters for several subsamples of the dwarfs.

%-------------------------------------------------

\section{Conclusions} \label{sec:conclusions}
We have used the TNG50 simulation, with its combination of high-resolution and large volume, to explore the cumulative SFHs of more than 15000 dwarf galaxies within the mass range of $\MSTAR=10^{7-10}$ at $z=0$ and encompassing a wide range of cosmological environments. We have studied and contrasted the dwarfs by subdividing them into centrals, satellites in hosts of mass $\MHOST=10^{12-14.3}\MSUN$ and dwarf group satellites in hosts of mass $\MHOST=10^{9.7-12}\MSUN$. While the dwarf group members have several properties that are intermediate between those of the centrals and satellites, such as colour and SFRs, their SFHs are on average similar to those of the centrals. Apart from examining the averaged normalized cumulative SFHs of various subsamples, we have also characterized the SFHs using the summary statistics $\tau_{10}$, $\tau_{50}$, $\tau_{90}$, the times when each galaxy had assembled 10, 50, and 90 per cent of its $z=0$ stellar mass. There is a large diversity in the cumulative SFHs of the dwarfs predicted by the TNG50 model, driven by several factors. In this paper we have focused on the effects of environment. Our findings are summarized as follows:

\begin{itemize}
    \item The cumulative SFHs of the dwarfs are, to zeroth-order, determined by their status as satellites or centrals, with the satellites having assembled 90 per cent of their stellar mass  $\sim7_{-5.5}^{+3.3}$~Gyr ago ($z\sim0.75$, on average and within the $10^{\rm th}-90^{\rm th}$ percentiles), whereas the centrals did so only $\sim1_{-0.5}^{+4.0}$~Gyr ago ($z\sim0.075$). For all the analyses of SFHs presented in this paper, the results for the dwarf group members are broadly consistent with those of the centrals.
    
    \item Satellite SFHs are strongly dependent on stellar mass, with low mass ($\MSTAR=10^{7-7.5}\MSUN$) satellites having assembled their stellar mass $\sim6$~Gyr earlier than the more massive satellites ($\MSTAR=10^{9.5-10}\MSUN$) in the same range of host masses. On the other hand, the central dwarf SFHs show little dependence on stellar mass.
    
    \item The median values of $\tau_{90}$ (in lookback time) decrease monotonically with increasing satellite mass, with some flattening seen at $\MSTAR\sim10^{9-10}\MSUN$, but with a significant scatter (quantified as half the difference between the $10^{\rm th}-90^{\rm th}$) of $\sim3.3-4.2$ Gyr that is approximately constant at all masses considered here. In the case of the centrals the values are nearly constant and with little scatter ($\sim0.75$ Gyr) at all higher stellar masses ($\MSTAR\sim10^{8-10}\MSUN$). The values of $\tau_{90}$ (in lookback time) are also consistently higher for satellites than centrals, implying that the satellites have assembled their stellar mass at earlier times and are largely quenched by $z=0$.
    
    \item In the case of the satellites, the cumulative SFHs are additionally dependent on host mass, such that dwarfs in more massive hosts assembled their stellar mass earlier. Overall, it is the ratio of stellar mass to host virial mass that best captures the variations of the satellites' cumulative SFHs. Furthermore, the satellites show secondary dependencies, even at fixed satellite and host mass, on position and accretion time, with satellites found closest to their host centres and those accreted earliest having built up their stellar mass earliest. These trends are well encapsulated by separating the satellite dwarfs into distinct regions in projected phase-space following \citet{Rhee2017}. Dwarfs found at small distances and with lower relative velocities, likely to be ancient infallers, have earlier SFHs compared to those at large distances and with higher relative velocities, likely to be first-infallers.
    
    \item The cumulative SFHs have at best a secondary dependence, negligible in the case of the centrals, on local environment defined here as the distance to the nearest massive ($\MSTAR>10^{10}\MSUN$) neighbour, with dwarfs that have nearer massive neighbours being the ones to build up their stellar mass earlier.
    
    \item The value of $\tau_{90}$ is often taken to be a proxy for the quenching time of galaxies; this assumption obviously does not hold for galaxies that are still star-forming at $z=0$. As for the galaxies that are quenched at $z=0$, our results show that while this is a fair assumption especially for galaxies that quenched early, the delay between the galaxies assembling 90 per cent of the stellar mass and quenching (defined as when the galaxy has an SFR of $<\rm{SFMS}-1$ dex) can be as large as 3 Gyr for those that quenched most recently.
    
    \item The shapes of the stacked cumulative SFHs are a product of the differently shaped individual SFHs of quenched and star-forming galaxies and the varying proportion of the two populations in any given subsample. This is also the case for any correlations between $\tau_{90}$ or $\tau_{50}$ and stellar mass. In fact, star-forming dwarfs have similarly shaped SFHs regardless of whether they are satellites or centrals, with little dependence on stellar mass. The SFHs of star-forming galaxies can be well parametrized by a two-phase model, the first characterized by a constant SFR and the second by a constant SFR. The quenched galaxies' SFHs are similarly well parametrized by a three-phase model, where the first two phases are identical to those for star-forming galaxies, and the third is characterized by an exponentially declining SFR.
    
    \item According to TNG50, there exists a subtle evolutionary and environmental effect whereby the satellite dwarfs in the most massive hosts at $z=0$ build up more stellar mass \emph{before} accretion in comparison to a control sample of central dwarfs that are mass-matched at accretion. This indicates an influence of the large-scale structure on future satellites, albeit admittedly mild, whereby the presence of a denser environment allows for higher SFRs at early times, before the dwarfs are eventually accreted onto their final hosts. This effect is weaker/non-existent for satellites in lower mass hosts.
    
\end{itemize}

In this study, we have shown that state-of-the-art galaxy formation models like TNG50 naturally return a great diversity in SFHs of dwarfs galaxies, both for centrals and satellites. A large contribution to such galaxy-to-galaxy variation is understood to be driven by a variety of factors, chiefly environmental processes, whose effects we have quantified in this paper. However, it should be noted that, when considering our results, the median cumulative SFHs are representative of the TNG50 dwarfs only in an average sense and that individual simulated SFHs can be significantly different from the mean or median, despite having similar $\tau_{50}$ values, especially in the case of satellites. This is particularly important when comparing the average outcomes of TNG50 to observational results of individual or smaller samples of objects, like those of the observed satellites of the MW or Local Volume galaxies. The comparisons that we have shown in this work are taken at face-value, without mocking either the observable or the sample selections to match the choices of available results from the literature. It remains to be determined if TNG50 contains analogs of the MW and/or Local Volume with populations of dwarfs which exhibit cumulative SFHs consistent with current observational constraints.

Finally, we have quantified the impact of various environmental factors (i.e. host mass, radial position with the host and accretion time), whose effects are known in determining several satellite galaxy properties. However, to our knowledge this is the first time that the effects of environments and environmental quenching have been cast in terms of how they shape the stacked cumulative SFHs, thanks to a large and diversified sample of simulated dwarfs. In fact, we have shown that the large diversity of SFHs, particularly when considering the stacked SFHs of different subsamples, is in large part the manifestation of varying proportions of star-forming vs. quenched galaxies in the given subsample, each of which exhibit markedly distinct cumulative SFHs shapes. Hence, results between mixed populations of galaxies should be compared with caution. Although central dwarfs are predicted by TNG50 to span a smaller range of cumulative SFHs compared to the satellites, it is nonetheless non-negligible. In an upcoming companion paper, we explore in detail the various factors that influence the central cumulative SFHs. Our results here and in the companion paper provide theoretical predictions for comparison to future, resolved stellar population observations from upcoming telescopes such as the JWST, the WFIRST/Roman, the LSST/Rubin and, further down the line, the ELT with MICADO.

%-------------------------------------------------

\section*{Acknowledgements}

The authors thank Anna Gallazzi and Marina Rejkuba for useful conversations and inputs. This work was funded by the Deutsche Forschungsgemeinschaft (DFG, German Research Foundation) -- Project-ID 138713538 -- SFB 881 (``The Milky Way System'', subproject C09). The TNG50 flagship simulation and its lower resolution counterparts were realised with compute time granted by the Gauss Centre for Super-computing (GCS) under GCS Large-Scale Project GCS-DWAR (2016; PIs Nelson/Pillepich). FM acknowledges support through the program ``Rita Levi Montalcini'' of the Italian Miur.

\section*{Data availability}
The codes used to produce the results presented in this work are available upon request from the corresponding author. Data for the TNG50 simulation series is publicly available as of February 1st, 2021, from the IllustrisTNG repository: \url{https://www.tng-project.org}.

%%%%%%%%%%%%%%%%%%%%%%%%%%%%%%%%%%%%%%%%%%%%%%%%%%



\bibliographystyle{mnras}
\bibliography{TNGDwarfSFHs} % if your bibtex file is called example.bib

%\begin{thebibliography}{99}
%\end{thebibliography}

%%%%%%%%%%%%%%%%%%%%%%%%%%%%%%%%%%%%%%%%%%%%%%%%%%

%%%%%%%%%%%%%%%%% APPENDICES %%%%%%%%%%%%%%%%%%%%%

\appendix

\section{Effects of numerical resolution on the derived cumulative SFHs} \label{sec:appResEffects}

\begin{figure*}
    \includegraphics[width=\linewidth]{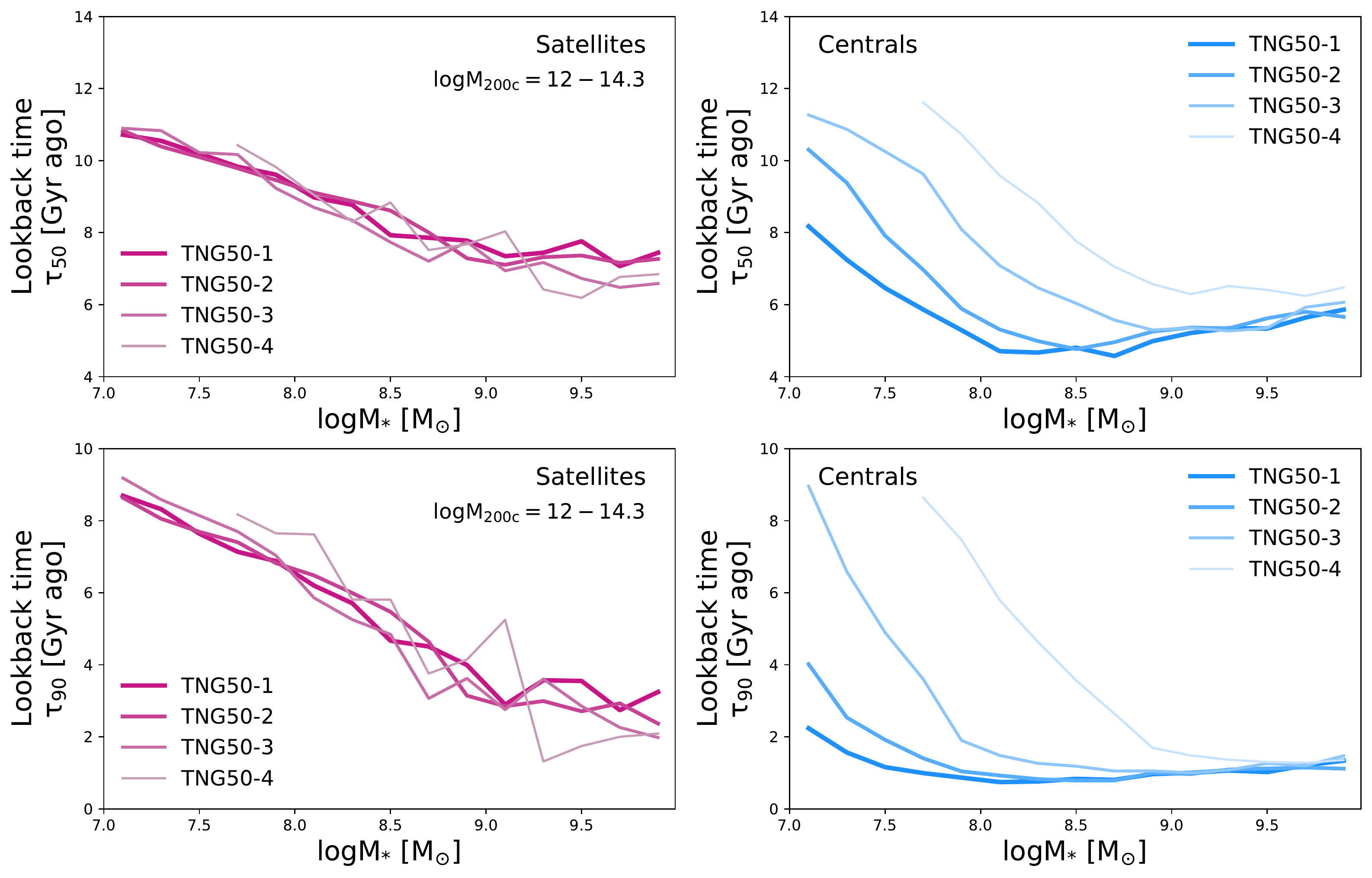}
    \caption{Median values of $\tau_{50}$ (top row) and $\tau_{90}$ (bottom row) as a function of stellar mass for satellites (left column) and centrals (right column), as in Fig.\ref{fig:t50t90AssemblyTimes}, from the fiducial highest resolution run of TNG50, i.e. TNG50-1 (thickest, darkest curves) to the lowest resolution run, TNG50-4 (thinnest, lightest curves). The results are well converged for all satellites, as well as for centrals with $\MSTAR=10^{8-10}\MSUN$. For centrals at masses of $\MSTAR=10^{7-7.5}\MSUN$ ($\MSTAR=10^{7.5-7}\MSUN$), the $\tau_{90}$ values are within 2 (1) Gyr of the results from the next lower resolution run. Similarly, the differences in $\tau_{50}$ values between consecutive resolutions are also within $\sim 2$ Gyr for $\MSTAR=10^{7-8}\MSUN$.} \label{fig:t50t90ResEffects}
\end{figure*}

\begin{figure*}
    \includegraphics[width=\linewidth]{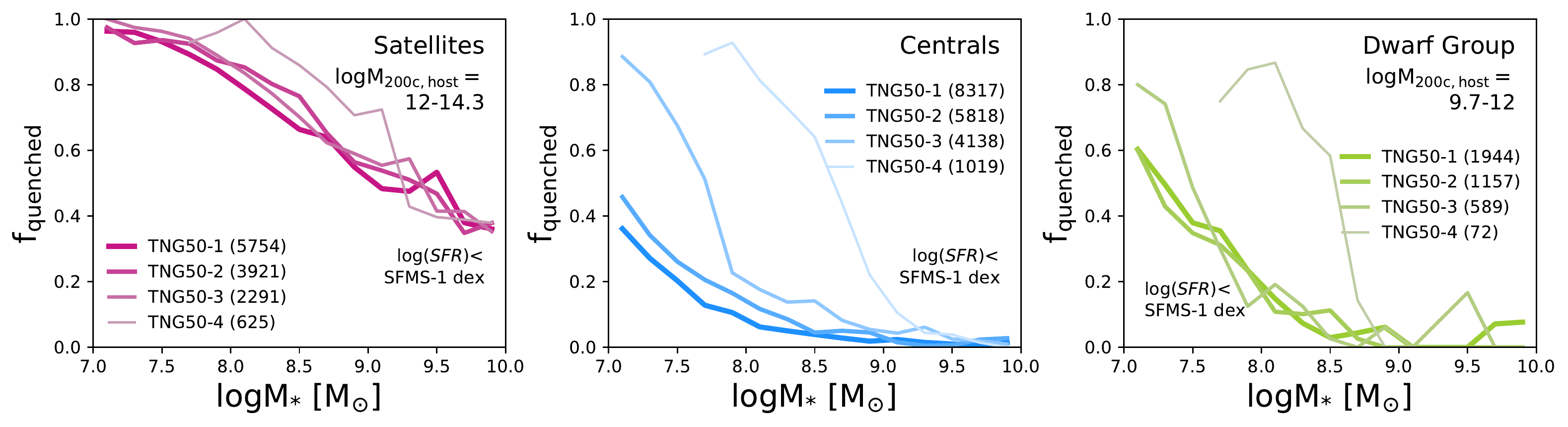}
    \caption{Quenched fractions as a function of stellar mass as in Fig. \ref{fig:sfVsQuenched} (a) for satellites (left), centrals (middle) and dwarf group members (right), from the fiducial highest resolution run of TNG50, i.e. TNG50-1 (thickest, darkest curves) to the lowest resolution run, TNG50-4 (thinnest, lightest curves). The quenched fractions are well converged for the satellites at all masses and for centrals and dwarf group members with $\MSTAR=10^{8-10}\MSUN$. At lower masses, the results are converging with the differences between TNG50-1 and TNG50-2 being at most 10 percentage points.} \label{fig:resEffectsQuenchedFracs}
\end{figure*}

To address the effects of numerical resolution on our findings, we discuss here a series of checks based on comparing the TNG50 results to those from its three lower resolution versions, called TNG50-2, TNG50-3 and TNG50-4, which are composed of resolution elements with target masses 8, 64, and 512 times larger than TNG50, with e.g. stellar particles of masses of $6.8\times10^{5}\MSUN$, $5.4\times10^{6}\MSUN$, $4.3\times10^{7}\MSUN$. These runs contain 11037, 7608 and 2291 dwarfs within the same mass range (note that in the case of TNG50-4, the actual mass range is $\MSTAR=10^{7.5-10}\MSUN$ as there is only one galaxy with $\MSTAR<10^{7.5}\MSUN$). However, we must exclude 1.3, 7.8 and 25.1 per cent of these samples, respectively, which cannot be traced back to $z=2$ as with the TNG50 sample. The remaining samples thus consist of 3921, 2291 and 625 satellites and 5818, 4138 and 1019 centrals. 

To quantify, within the TNG model, whether and to what level the results presented in the paper with TNG50 are converged, we have studied the cumulative SFHs, the distributions of $\tau_{10}$, $\tau_{50}$ and $\tau_{90}$, quenched fractions, and the dependence on the various environmental factors, all in bins of stellar mass, in all the simulations of the TNG50 series. Here we explicitly show the results for the stellar assembly times (Fig.~\ref{fig:t50t90ResEffects}) and for the quenched fractions (Fig.~\ref{fig:resEffectsQuenchedFracs} -- see also the figures in the Appendix of \citealt{Donnari2020a, Donnari2020b} for similar studies extending to larger galaxy and host masses and higher redshifts, with TNG100 and TNG300). We also report on other findings without necessarily showing their corresponding plots, for brevity.

In Figs.~\ref{fig:t50t90ResEffects} and \ref{fig:resEffectsQuenchedFracs}, we reproduce the median $\tau_{50}$ (top row) and $\tau_{90}$ (bottom row) values from Fig. \ref{fig:t50t90AssemblyTimes} and the quenched fractions from Fig.~\ref{fig:sfVsQuenched} (a), as a function of stellar mass, and show the same results at the three lower resolutions of TNG50. In the case of the satellites, both the quenched fractions and median values of $\tau_{50}$ and $\tau_{90}$ as a function of stellar mass are in good agreement for the three highest-resolution runs. Overall, we find that the impact of a massive host halo, which itself is likely to be well resolved, dominates any resolution effects that may have been present in the average properties of the satellites.

For the centrals, on the other hand, the values of $\tau_{50}$ and $\tau_{90}$, as well as the quenched fractions, are higher at lower resolutions for $\MSTAR\lesssim10^{8.5}\MSUN$, with the largest differences seen at the lowest masses. However, although the results are not fully converged, they are \emph{converging}; this is the case of $\tau_{90}$ at all masses and of $\tau_{50}$ for stellar masses larger than $\MSTAR\gtrsim10^{7.5}\MSUN$. In other words, the differences between TNG50 and TNG50-2 are smaller than those between TNG50-2 and TNG50-3, which in turn are smaller than those between TNG50-3 and TNG50-4. The median values of $\tau_{90}$ as a function of stellar mass are converged to within 2~Gyr for the lower mass galaxies ($\MSTAR\lesssim10^{7.2}\MSUN$), within 1~Gyr for centrals with $\MSTAR\approx^{7.2-7.8}\MSUN$ and well converged at higher masses. Similarly, although we do not see the median values of $\tau_{50}$ to be converging at the lowest mass end, the differences between consecutive resolution levels are smaller than 2~Gyr and are within a few 100 Myr for $\MSTAR \gtrsim 10^{8}\MSUN$ across all the three higher resolution runs.

We have also checked the differences in cumulative SFHs across the different resolutions (not shown for brevity, but reflecting the findings for the stellar assembly times shown in Fig.~\ref{fig:t50t90ResEffects}). The cumulative SFHs of the satellites are well converged at all masses in our TNG50 sample, i.e, in the case of the satellites, we find that with the exception of TNG50-4 (the lowest resolution run, even inferior to that of TNG300 which is unsuitable for studying such low-mass galaxies as studied in this analysis) most of the results shown in the main text of the paper are remarkably robust with resolution, within the values probed here, and that the results of TNG50 appear converged to any relevant level of accuracy.

As for the centrals, we do see some effects of resolution for dwarfs with $\MSTAR\lesssim10^{8.5}\MSUN$, namely that the average cumulative SFHs in bins of stellar mass tend towards an earlier build up of stellar mass and towards earlier quenching times with progressively worse resolution. However, the TNG50-2 results are indistinguishable from those of TNG50-1 at masses of $\MSTAR>10^{8}\MSUN$, corresponding to a minimum of $\sim 150$ stellar particles in TNG50-2; at lower masses the largest difference in cumulative SFHs is $\sim 0.15$ in $\MSTAR(z)/\MSTAR(z=0)$. Similarly, the TNG50-3 results are identical to the TNG50-2 results at masses of $\MSTAR>10^{8.5}\MSUN$, corresponding to a minimum of $\sim 60$ stellar particles in TNG50-3; the maximum differences at lower masses is $\sim 0.3$. Therefore, we can conclude that the results for TNG50 centrals are well converged to any relevant level of accuracy except for galaxies with $\MSTAR\lesssim10^{7-7.5}\MSUN$, for whom some quantitative, albeit small, systematic effects need to be accounted for.

\section{Best-fit model parameters for cumulative SFHs} \label{sec:appModelFits}
In this section, we provide the best-fitting parameters for the piecewise-defined models introduced in Section~\ref{sec:discModel}, describing the SFHs of dwarfs predicted by TNG50. Table~\ref{tab:fitParsSF} provides the values for $\alpha$ and $t_{1}$ for star-forming centrals in bins of stellar mass and for star-forming satellites in bins of host mass, stellar mass, and further separated by the phase-space regions of \citet{Rhee2017}. Table \ref{tab:fitParsQuenched} provides the values of $\alpha$, $t_{1}$, $t_{2}$ and $\tau$ for the corresponding quenched dwarfs. In all cases, we only calculate parameters for subsamples containing at least 3 galaxies.

\begin{table}
    \centering
    \caption{Best fit parameters for star-forming centrals in bins of stellar mass and for star-forming satellites in bins of stellar mass and host mass and further subdivided into the phase-space regions of \citet{Rhee2017} for the model described in Section~\ref{sec:discModel} and defined by equations (\ref{eq:modelSF}) and (\ref{eq:m0SF}). Parameters are only calculated for subsamples containing at least three galaxies.} \label{tab:fitParsSF}
    \begin{tabular}{c|c|c|c}
        \hline
        $\log{M_{*}}$ & Phase- & $\alpha$ & $t_{1}$ \\
        & space & & \\
        $[\MSUN]$ & region & [Gyr$^{-1}$] & [Gyr] \\
        \hline
        \multicolumn{4}{c}{Star-forming centrals} \\
        \hline
        7.0-8.0 & - & 0.242 & 6.81 \\
8.0-9.0 & - & 0.259 & 8.23 \\
9.0-10.0 & - & 0.474 & 5.13 \\

        \hline
        \multicolumn{4}{c}{Star-forming satellites with $\MHOST=10^{12-13}\MSUN$} \\
        \hline
        7.0-8.0 & A & 0.316 & 5.93 \\
& C & 0.453 & 3.62 \\
& D & 0.308 & 6.62 \\
& E & 0.308 & 5.92 \\
8.0-9.0 & A & 0.387 & 5.79 \\
& C & 0.353 & 5.75 \\
& D & 0.572 & 3.92 \\
& E & 0.343 & 6.10 \\
9.0-10.0 & A & 0.717 & 3.97 \\
& C & 0.845 & 3.27 \\
& D & 0.635 & 4.27 \\
& E & 0.576 & 4.60 \\

        \hline
        \multicolumn{4}{c}{Star-forming satellites with $\MHOST=10^{13-14}\MSUN$} \\
        \hline
        7.0-8.0 & A & 0.320 & 6.53 \\
& C & 0.278 & 7.30 \\
& D & 0.886 & 2.39 \\
& E & 1.832 & 0.25 \\
8.0-9.0 & A & 0.428 & 4.68 \\
& C & 0.503 & 4.24 \\
& D & 0.546 & 4.30 \\
& E & 0.422 & 5.52 \\
9.0-10.0 & A & 0.910 & 3.08 \\
& C & 1.406 & 2.61 \\
& D & 0.439 & 5.57 \\
& E & 0.712 & 3.59 \\

        \hline
        \multicolumn{4}{c}{Star-forming satellites with $\MHOST=10^{14-14.3}\MSUN$} \\
        \hline
        7.0-8.0 & D & 0.303 & 6.19 \\
8.0-9.0 & A & 0.335 & 6.36 \\
& C & 1.181 & 2.19 \\
& D & 0.358 & 6.40 \\
9.0-10.0 & A & 3.922 & 1.30 \\
& C & 2.731 & 1.55 \\

        \hline
    \end{tabular}
\end{table}

\begin{table}
    \centering
    \caption{Best fit parameters for quenched and green-valley centrals in bins of stellar mass and satellites in bins of stellar mass and host mass and further subdivided into the phase-space regions of \citet{Rhee2017} for the model described in Section~\ref{sec:discModel} and defined by equations (\ref{eq:modelQuenched}) and (\ref{eq:m0Quenched}). Parameters are only calculated for subsamples containing at least three galaxies.} \label{tab:fitParsQuenched}
    \begin{tabular}{c|c|c|c|c|c}
        \hline
        $\log{M_{*}}$ & Phase & $\alpha$ & $t_{1}$ & $t_{2}$ & $\tau$ \\
        & -space & & & & \\
        $[\MSUN]$ & region & [Gyr$^{-1}$] & [Gyr] & [Gyr] & [Gyr] \\
        \hline
        \multicolumn{6}{c}{Quenched centrals} \\
        \hline
        7.0-8.0 & - & 3.396 & 0.91 & 3.45 & 4.13 \\
8.0-9.0 & - & 1.384 & 1.98 & 9.82 & 1.22 \\
9.0-10.0 & - & 0.856 & 3.16 & 10.12 & 3.08 \\

        \hline
        \multicolumn{6}{c}{Quenched satellites with $\MHOST=10^{12-13}\MSUN$} \\
        \hline
        7.0-8.0 & A & 2.063 & 1.35 & 7.16 & 1.19 \\
& B & 1.566 & 1.69 & 6.30 & 0.35 \\
& C & 3.393 & 1.04 & 5.84 & 0.80 \\
& D & 1.842 & 1.58 & 5.98 & 0.79 \\
& E & 1.970 & 1.64 & 5.62 & 0.74 \\
8.0-9.0 & A & 1.222 & 2.36 & 10.54 & 1.21 \\
& B & 1.899 & 1.87 & 8.43 & 0.71 \\
& C & 1.624 & 1.87 & 7.25 & 1.64 \\
& D & 1.387 & 2.18 & 7.20 & 1.74 \\
& E & 1.259 & 2.52 & 7.52 & 1.11 \\
9.0-10.0 & A & 1.544 & 2.18 & 9.11 & 3.27 \\
& C & 1.039 & 2.63 & 7.00 & 3.16 \\
& D & 0.859 & 3.12 & 8.39 & 2.62 \\
& E & 1.076 & 3.00 & 8.06 & 1.95 \\

        \hline
        \multicolumn{6}{c}{Quenched satellites with $\MHOST=10^{13-14}\MSUN$} \\
        \hline
        7.0-8.0 & A & 2.999 & 1.13 & 5.72 & 1.05 \\
& B & 3.375 & 1.05 & 6.09 & 0.29 \\
& C & 2.203 & 1.40 & 5.34 & 0.63 \\
& D & 2.269 & 1.39 & 5.17 & 0.70 \\
& E & 1.857 & 1.68 & 4.43 & 0.38 \\
8.0-9.0 & A & 1.636 & 1.89 & 7.92 & 1.41 \\
& B & 0.561 & 4.11 & 8.07 & 2.05 \\
& C & 1.201 & 2.37 & 6.99 & 0.80 \\
& D & 1.287 & 2.33 & 7.56 & 0.69 \\
& E & 1.348 & 2.38 & 5.30 & 0.49 \\
9.0-10.0 & A & 1.163 & 2.58 & 8.13 & 2.07 \\
& C & 0.995 & 3.06 & 8.01 & 1.34 \\
& D & 0.762 & 3.74 & 8.35 & 1.16 \\
& E & 1.035 & 3.36 & 5.71 & 0.81 \\

        \hline
        \multicolumn{6}{c}{Quenched satellites with $\MHOST=10^{14-14.3}\MSUN$} \\
        \hline
        7.0-8.0 & A & 6.297 & 0.71 & 7.22 & 0.96 \\
& B & 1.559 & 1.85 & 4.08 & 1.18 \\
& C & 2.703 & 1.25 & 4.47 & 0.87 \\
& D & 2.410 & 1.39 & 4.50 & 0.62 \\
& E & 1.996 & 1.66 & 3.82 & 0.29 \\
8.0-9.0 & A & 2.746 & 1.45 & 10.01 & 0.01 \\
& B & 4.239 & 1.16 & 4.00 & 0.67 \\
& C & 1.503 & 2.11 & 8.06 & 0.53 \\
& D & 2.452 & 1.52 & 5.96 & 0.72 \\
& E & 1.440 & 2.32 & 4.06 & 0.39 \\
9.0-10.0 & A & 1.326 & 2.63 & 8.88 & 0.68 \\
& C & 1.193 & 2.79 & 7.92 & 0.41 \\
& D & 1.513 & 2.30 & 5.94 & 1.11 \\
& E & 1.095 & 2.79 & 4.68 & 0.67 \\

        \hline
    \end{tabular}
\end{table}

%%%%%%%%%%%%%%%%%%%%%%%%%%%%%%%%%%%%%%%%%%%%%%%%%%

% Don't change these lines
\bsp	% typesetting comment
\label{lastpage}
\end{document}